\documentclass[11pt,a4paper]{article}
\usepackage{amsmath}
\usepackage{amsfonts}
\usepackage{color}
\usepackage{graphicx}
\usepackage{dcolumn}
\usepackage{bm}
\usepackage[
bookmarks,bookmarksnumbered,colorlinks=true,anchorcolor=blue,
linkcolor=blue,urlcolor=blue,citecolor=blue,
breaklinks=true]{hyperref}
\usepackage{geometry}
\usepackage{authblk}
\usepackage{amssymb}
\usepackage[utf8]{inputenc}
\usepackage[numbers,sort&compress]{natbib}
\usepackage{ulem}
\usepackage{tikz}

\numberwithin{equation}{section}   

\def\s{{\rm s}}
\def\r{{\rm r}}
\def\a{{\rm a}}

\def\tomega{{\tilde \omega}}

\def\tk{{\tilde k}}

\date{\today}

\begin{document}

\title{\bf Holographic derivation of effective action for a dissipative neutral fluid}


\author[1]{Yanyan Bu \thanks{yybu@hit.edu.cn}}
\author[1]{Xiyang Sun \thanks{xysun@stu.hit.edu.cn (correspondence author)}}

\affil[1]{\it School of Physics, Harbin Institute of Technology, Harbin 150001, China}

\maketitle

\begin{abstract}
Using a holographic prescription for the Schwinger-Keldysh closed time path, we derive the effective action for a dissipative neutral fluid holographically described by the Einstein gravity in an asymptotic AdS spacetime. In the saddle point approximation for the dual gravity, the goal is achieved by solving the double Dirichlet problem for the linearized gravitational field living in a complexified static AdS black brane background. We adopt a partially on-shell scheme for solving the bulk dynamics, which is equivalent to ``integrating out'' the gapped modes in the boundary field theory. The boundary effective action in the fluid spacetime, identified as the partially on-shell bulk action, is computed to first order in boundary derivative and to cubic order in AdS boundary data. The boundary effective action, rewritten in the physical spacetime, successfully reproduces various results known in the framework of classical hydrodynamics, confirming our holographic derivation.
\end{abstract}

\newpage

\tableofcontents

\allowdisplaybreaks

\flushbottom

\section{Introduction}

The AdS/CFT correspondence or holographic duality \cite{Maldacena:1997re,Gubser:1998bc,Witten:1998qj} has provided us with a tractable tool for exploring the properties of strongly coupled systems at finite temperature. In the large $N_c$ limit, the quantum dynamics of a many-body system living on the conformal boundary of an AdS spacetime is mapped to the classical dynamics of certain fields propagating in the entire bulk of the AdS space. Perhaps a most insightful application of the holographic duality is to study the transport properties of strongly coupled field theories \cite{Policastro:2001yc,Policastro:2002se,Policastro:2002tn,Son:2007vk}, which suggests a deep link between the quantum many-body physics and the black hole physics \cite{Son:2007vk}. The holographic duality predicts a universal value for shear viscosity over entropy density ratio for a large class of strongly coupled field theories \cite{Policastro:2001yc,Kovtun:2004de,Buchel:2003tz,Iqbal:2008by,Cai:2008ph}. Intriguingly, the holographic prediction for this ratio is quite close to those of two strongly correlated quantum liquids created in the labs: the quark-gluon plasma and the unitary Fermi gas \cite{Adams:2012th}.

Indeed, there is a basic assumption behind the holographic studies mentioned above: the long-time long-distance limit of an interacting quantum field theory at finite temperature is effectively described by hydrodynamics \cite{landau,Kovtun:2012rj}. Then, the sophisticated dynamics at microscopic scale is replaced by hydrodynamic equation of motion (conservation law) at large scale
\begin{align}
\nabla_\mu T_{\rm hydro}^{\mu\nu} = 0,  \label{hydro_eom}
\end{align}
which is supplemented with a hydrodynamic constitutive relation
\begin{align}
T_{\rm hydro}^{\mu\nu} = (\epsilon + P) u^\mu u^\nu + P g_{\r}^{\mu\nu} - \eta_0 \sigma^{\mu\nu} + \cdots. \label{hydro_const_relation}
\end{align}
Here, $\epsilon$ and $P$ are the energy density and pressure, $u^\mu$ is the fluid velocity (a collective variable), $g_{\r}^{\mu\nu}$ is a curved background metric, $\sigma^{\mu\nu}$ is the shear tensor, and $\eta_0$ is the shear viscosity characterizing off-equilibrium properties of the system. With the assumption \eqref{hydro_eom} and \eqref{hydro_const_relation}, the shear viscosity $\eta_0$ is related to retarded two-point correlation function of stress tensor $T_{\rm hydro}^{\mu\nu}$ (the Kubo formula), which can be calculated holographically via solving the linearized Einstein equations in the AdS black brane \cite{Son:2002sd}.

Later on, the connection between the classical hydrodynamics and the AdS gravity was made more transparent, accumulated in the fluid-gravity correspondence \cite{Bhattacharyya:2008jc,Rangamani:2009xk}, which establishes a one-to-one correspondence between the classical hydrodynamics \eqref{hydro_eom}-\eqref{hydro_const_relation} and the solutions to the Einstein equations in the AdS space. In principle, the fluid-gravity correspondence allows us to systematically derive the hydrodynamic constitutive relation \eqref{hydro_const_relation} order-by-order in terms of the hydrodynamic derivative expansion.

\begin{figure}[htbp!]
	\centering
	\begin{tikzpicture}[]
	\draw[very thick] (-5,-0.2)--(5,-0.2);
	\node[above] at   (0,0.2) {$U(t_f,t_i)$};
	\draw[very thick] ( 4.98,-0.2)--(4.98, 0.2);
	\node[above] at   (5,0.2) {$t_f$};
	\draw[very thick] (-5, 0.2)--(5, 0.2);
	\node[below] at   (0,-0.2) {$U^\dagger(t_f,t_i)$};
	\draw[ ->,very thick] (-1,0.2)--(0.14,0.2);
	\draw[ <-,very thick] (0,-0.2)--(1,-0.2);
	\draw[fill]  (-5,-0.2) circle [radius=0.1];
	\draw[fill]  (-5,0.2) circle [radius=0.1];
	\node[left]  at (-5,0) {$\rho_0$};
	\node[above] at (-5.3,0.2) {$t_i$};
	\end{tikzpicture}
\caption{The SK closed time path with an initial state $\rho_0$. Here, $U(t_f,t_i)$ is the time-evolution operator from initial time $t_i$ to final time $t_f$.} \label{SK contour}
\end{figure}
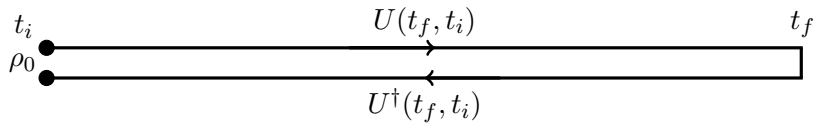

However, the holographic approach to hydrodynamics \cite{Policastro:2001yc,Policastro:2002se,Policastro:2002tn,Son:2007vk,Bhattacharyya:2008jc,Rangamani:2009xk} ignores the thermal fluctuations of dynamical variables as required by the fluctuation-dissipation theorem. This drawback is related to the ingoing wave condition \cite{Son:2002sd} imposed for a bulk field near the event horizon, which was proven to be sufficient for addressing dissipations \cite{Son:2002sd}. From the perspective of the black hole physics, the outgoing wave condition is also possible for a bulk field and would intuitively correspond to the thermal fluctuation \cite{Herzog:2002pc,Son:2009vu} for the boundary system. Ideally, a bulk field consistently containing both ingoing and outgoing modes presumably gives rise to the full set of the Schwinger-Keldysh (SK) correlators \cite{Herzog:2002pc}, including retarded, advanced, and symmetric ones, etc. Actually, for a system in a mixed state denoted by a density matrix $\rho_0$, an ideal framework for addressing dissipations and fluctuations is the SK closed time path shown in Figure \ref{SK contour}.
Basically, within the SK formalism, each dynamical variable $\psi$ of the system gets doubled, $\psi \to (\psi_1, \psi_2)$, where the subscripts 1 and 2 denote variables on the upper and lower branches of the SK contour in Figure \ref{SK contour}. When the initial state is a thermal one, holographic prescriptions were proposed in \cite{Herzog:2002pc,Skenderis:2008dh,Skenderis:2008dg,Glorioso:2018mmw} for the SK closed time path of Figure \ref{SK contour}.

The holographic approach to hydrodynamics has inspired deep investigation on the theoretical foundation of fluid dynamics. Among others, an action principle has been recently formulated for dissipative hydrodynamics in \cite{Crossley:2015evo,Glorioso:2017fpd,
Haehl:2015uoc,Haehl:2018lcu}\footnote{For early attempts on this subject, see e.g., \cite{Dubovsky:2011sj,Endlich:2012vt,Grozdanov:2013dba,Kovtun:2014hpa,Haehl:2015foa,
Montenegro:2016gjq}. Further exploration on the formal aspects of the hydrodynamic effective field theory can be found in e.g., \cite{Glorioso:2016gsa,Gao:2018bxz,Jensen:2017kzi}.} (see \cite{Liu:2018kfw} for a nice review) by virtue of the SK formalism. Basically, a local effective action can be written down for a dissipative fluid, in which the dynamical variables correspond to the SK-doubled fluid velocity and temperature fields. A main motivation behind such a study is to resolve the shortcomings of the classical hydrodynamics \cite{Crossley:2015evo,Liu:2018kfw}. The classical hydrodynamics is essentially phenomenological: in order to ensure such a framework to work well, several constraints have to be imposed by hand, such as the second law of thermodynamics, the fluctuation-dissipation relations, etc. However, an action principle for hydrodynamics has at least two advantages. First, all the ingredients of classical hydrodynamics, including \eqref{hydro_eom}-\eqref{hydro_const_relation} as well as various phenomenological constraints, are integrated into a symmetry-based effective action. Second, the hydrodynamic effective field theory (EFT) provides a systematic treatment over the thermal fluctuation of dynamical variable, which is usually modelled by Gaussian stochastic force \cite{Kovtun:2012rj} replacing the right-handed side of the hydrodynamic equation \eqref{hydro_eom}.

It is then of interest to understand the hydrodynamic EFT from the holographic perspective. Previous attempts of deriving the hydrodynamic effective action from the AdS gravity can be found in \cite{Nickel:2010pr,Crossley:2015tka,deBoer:2015ija}. However, both programs \cite{Crossley:2015tka,deBoer:2015ija} had run into problems for the gravity metric near the horizon, which was avoided by a non-dissipative horizon condition. In recent years, the holographic SK technique of \cite{Glorioso:2018mmw} has been proven very efficient in deriving the effective action for certain boundary system at finite temperature, see, e.g., \cite{Glorioso:2018mmw,deBoer:2018qqm,Chakrabarty:2019aeu,Bu:2020jfo,Bu:2021clf,
Bu:2021jlp,Bu:2022esd,Bu:2022oty,Baggioli:2023tlc,Bu:2024oyz,Liu:2024tqe,Baggioli:2024zfq,Bu:2024fhz} for recent developments. However, these studies\footnote{While recent works \cite{He:2021jna,He:2022jnc,He:2022deg} considered the linearized AdS gravity using the prescription \cite{Glorioso:2018mmw}, we understand that they focused on the Wilsonian influence functional rather than on the off-shell effective action.} focus on the dynamics of matter fields in a complexified AdS black brane background and does not touch on the dynamics of bulk gravity itself. This amounts to neglecting the variations of fluid velocity and temperature in the boundary theory.

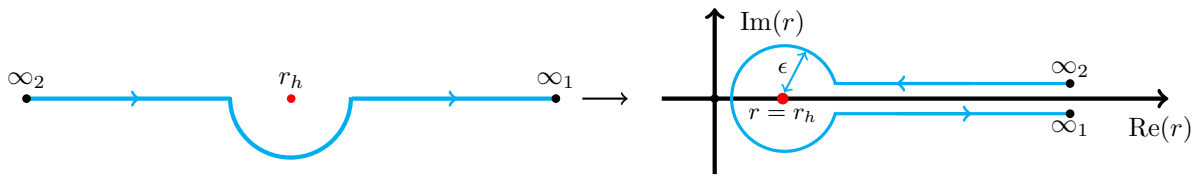
\begin{figure} [htbp!]
		\centering
		\begin{tikzpicture}[]
		
		\draw[cyan, ultra thick] (-5.5,0)--(-2.8,0);
		\draw[cyan, ultra thick] (-1.2,0)--(1.5,0);
		\draw[cyan, ultra thick] (-2.81,0.019) arc (-180:0:0.8);
		\draw[cyan, ->,very thick] (-1,0)--(0.2,0);
		\draw[cyan, ->,very thick] (-4.2,0)--(-4,0);
		\draw[fill] (-5.5,0) circle [radius=0.05];
		\node[above] at (-5.5,0) {\small $\infty_2$};
		\draw[fill] (1.5,0) circle [radius=0.05];
		\node[above] at (1.5,0) {\small $\infty_1$};
		\draw[fill,red] (-2,0) circle [radius=0.05];
		\node[above] at (-2,0) {\small $r_h$};				
		
		\draw[-to,thick] (1.85,0)--(2.45,0);

		\node[below] at (9.5,-0.1) {\small Re$(r)$};
		\draw[->,ultra thick] (2.9,0)--(9.6,0);
		\node[right] at (3.8,1) {\small Im$(r)$};
		\draw[->,ultra thick] (3.6,-1)--(3.6,1.2);
		\draw[cyan, very thick] (5.2,-0.18) arc (165:-165:-0.7);
		\draw[cyan, very thick] (5.2,0.2)--(8.3,0.2);
		\draw[cyan, very thick] (5.2,-0.2)--(8.3,-0.2);
		\draw[cyan, ->,very thick] (6,-0.2)--(7,-0.2);
		\draw[cyan, <-,very thick] (6,0.2)--(7,0.2);
		\draw[fill] (3.6,0) circle [radius=0.05];
		\draw[fill] (8.3,0.2) circle [radius=0.05];
		\node[below] at (8.3,-0.15) {\small $\infty_1 $};
		\draw[fill] (8.3,-0.2) circle [radius=0.05];
		\node[above] at (8.3,0.15) {\small $\infty_2 $};
		\draw[fill , red ] (4.5,0) circle [radius=0.07];
		\node[below] at (4.5,0) {\footnotesize $r=r_h $};
		\draw[cyan, thick,<->] (4.52,0.08)--(4.8,0.62);
		\node[above] at (4.5, 0.2) {\footnotesize $\epsilon$};
		\end{tikzpicture}
		\caption{A holographic prescription for the SK closed time path of Figure \ref{SK contour}. Left: complexified double AdS \cite{Crossley:2015tka}; Right: the holographic SK contour \cite{Glorioso:2018mmw}. Notice that the two horizontal legs in the right panel indeed overlap with the real axis.} \label{holographic_SK_contour}
\end{figure}

In this work, we use the holographic prescription of \cite{Glorioso:2018mmw} and derive the SK effective action for a dissipative neutral fluid. We will take a complexified static AdS black brane background (see Figure \ref{holographic_SK_contour}), which is dual to the SK closed time path with an initial state described by a thermal density matrix.
Then, we will work out the bulk metric perturbation on top of the static AdS black brane, which amounts to introducing variations as well as fluctuations for the fluid velocity and temperature on the boundary. This treatment is more akin to the holographic approach to hydrodynamics \cite{Policastro:2001yc,Policastro:2002se,Policastro:2002tn,Son:2007vk,deBoer:2015ija}, whilst the latter studies rely on single copy AdS.

The rest of this paper will be structured as follows. In section \ref{hydro_EFT_review} we will review the formulation of hydrodynamic EFT from dual perspectives: field theory versus holography. We have made refinements over the following points: 1) the expansion of EFT's building blocks in the linear regime; 2) the holographic program towards the boundary EFT action. Section \ref{holo_EFT} contains the main results of this work. In this section we will present the holographic derivation of the effective action for a dissipative neutral fluid. Meanwhile, we will confirm our results by recovering, from the holographic effective action, the hydrodynamic modes, the hydrodynamic constitutive relation (i.e., one-point function of stress tensor) and the stochastic version of \eqref{hydro_eom}, and the full set of two-point correlators of the stress tensor. In section \ref{summary} we present a brief summary and discuss some open questions. In appendix \ref{cal_detail} we record details for the generating functional obtained from the holographic effective action.

\section{Hydrodynamic effective field theory} \label{hydro_EFT_review}

In this section we review the formulation of an action principle for dissipative fluid dynamics. We will discuss such a program from dual perspectives: quantum field theory versus holographic duality. The emphasis will be on the general idea of the Wilsonian renormalization group (RG) applied to a quantum many-body system at finite temperature, and its implementation for a holographic system. While the discussion of this section will be found abstract, the main lesson can be summarized as two aspects (similar to the U(1) case):

On the field theory side, the building block for formulating the hydrodynamic EFT is $\mathcal G_{ab}(X^\mu(\sigma))$ to be introduced in \eqref{diffeo_transform_promotion}, which encodes both the external metric and the dynamical variable associated with the conserved energy and momentum. Moreover, the hydrodynamic EFT shall obey a set of proposed symmetries, see \eqref{normal_condition}-\eqref{dynamical_KMS}.

From the gravity perspective, the building block $\mathcal G_{ab}(X^\mu(\sigma))$ emerges via the boundary condition of the bulk metric at the AdS boundary, see \eqref{bdy_data} and \eqref{sigma_coord}. In addition, in order to ensure $X^\mu(\sigma)$ to be dynamical, we shall use a partially on-shell prescription when solving the bulk dynamics. Within this prescription, the dynamical equations are solved to obtain the bulk metric, while the constraint equations shall be relaxed.

\subsection{The perspective from field theory} \label{idea_field_theory}

Here, we present the basic aspects of hydrodynamic EFT from modern quantum field theory consideration. We will mainly follow the relevant discussions of \cite{Crossley:2015evo} (see also the review \cite{Liu:2018kfw}).

Imagine a quantum many-body system at finite temperature, whose microscopic description may involve a Lagrangian density $\mathcal{L}_0$, a local functional of certain microscopic degrees of freedom denoted by $\psi$. In order to study the real-time evolution of a macroscopic state, it is more convenient to put the system in the SK closed time path (see Figure \ref{SK contour}), so that the variable for each degree of freedom gets doubled $\psi \to \psi_{1,2}$.

Momentarily, we will not bother to make the SK indices in $\psi_{1,2}$ explicit. Moreover, we assume that the only conserved charges of the system are energy and momentum, cumulated in the energy-momentum tensor $T^{\mu\nu}$. The Noether theorem offers a way of constructing $T^{\mu\nu}$ in terms of the microscopic field $\psi$. Nevertheless, a more powerful treatment is to couple the system to an external (background) metric $g_{\mu \nu}(x)$ (usually via a minimal coupling scheme)
\begin{align}
\mathcal L_0 [\psi(x)] \to \mathcal L[\psi(x), g_{\mu\nu}(x)]
\end{align}
where the external metric $g_{\mu\nu}(x)$ is specified via the following line element
\begin{align}
dl_0^2 = g_{\mu\nu}(x) dx^\mu dx^\nu \label{dl02}
\end{align}
Now, we introduce the generating functional (or the partition function) of the system,
\begin{align}
Z[g_{\mu\nu}(x)] = \int_{\rho_0} [D \psi] e^{i \int d^4x \, \mathcal L[\psi,\, g_{\mu\nu}]}, \label{Z_micro}
\end{align}
where $\rho_0$ is the initial state of the system.

The conservation law of energy and momentum, $\nabla_\mu T^{\mu\nu} = 0$, can be guaranteed by imposing that $Z$ in \eqref{Z_micro} is invariant under the diffeomorphism transformation of the background metric
\begin{align}
Z[\bar g_{\alpha\beta}(\bar x)] = Z[g_{\mu\nu}(x)]
\end{align}
where
\begin{align}
& dl_0^2 = \bar g_{\alpha\beta}(\bar x) d\bar x^\alpha d\bar x^\beta = g_{\mu\nu}(x) dx^\mu dx^\nu \nonumber \\
& \Rightarrow \bar g_{\alpha\beta}(\bar x) = g_{\mu\nu}(x(\bar x)) \frac{\partial x^\mu} {\partial \bar x^\alpha} \frac{\partial x^\nu}{\partial \bar x^\beta} \label{diffeo_trans}
\end{align}
Physically, there is no essential difference between the two coordinate systems $\{\bar x^\alpha, \bar g_{\alpha\beta}(\bar x)\}$ and $\{x^\mu, g_{\mu\nu}(x)\}$. The Green's functions of the stress tensor are essentially functional derivatives of $Z$ with respect to the external metric $g_{\mu\nu}(x)$. In the hydrodynamic regime, it can be shown that \cite{Policastro:2002tn} the only singularities of the Green's functions for $T^{\mu\nu}$ are simple poles, representing the shear wave and sound wave. This fact immediately implies that $Z$ cannot be a local functional of the external metric $g_{\mu\nu}$ \cite{Crossley:2015evo}. Indeed, this reflects the fact that \cite{Crossley:2015evo} in \eqref{Z_micro} those gapless modes have been integrated out towards obtaining $Z$.

However, if one can identity the gapless modes and could perform an ``integrate-in'' procedure, one will be able to re-express the partition function $Z$ as a path integral over those gapless modes weighted with a ``local'' effective action. Here comes the idea of \cite{Crossley:2015evo}: the diffeomorphism transformation parameter in \eqref{diffeo_trans} is promoted to a dynamical field and is identified as the suitable parametrization of the gapless modes. Notice that the transformation $\bar x^\mu \to x^\mu$ in \eqref{diffeo_trans} can be thought of as generated by the diffeomorphism transformation parameter $x^\mu$, which becomes more obvious for an infinitesimal coordinate transformation. Eventually, by analogy with the situation of the charge diffusion \cite{Crossley:2015evo}, we promote the spacetime coordinate $x^\mu$ (in which the external metric is defined) into the dynamical field associated with the conserved energy and momentum
\begin{align}
x^\mu \to X^\mu(\sigma^a),
\end{align}
where $\sigma^a$ is an emergent coordinate so that $X^\mu$ is a dynamical field. In \cite{Crossley:2015evo}, the spacetime spanned by $\sigma^a$ is referred to as the fluid spacetime: the spatial part $\sigma^i$ labels a fluid element, and the time component $\sigma^0$ serves as an internal clock carried by a fluid element. Indeed, $X^\mu(\sigma^a)$ corresponds to the trajectory of the fluid element labeled by $\sigma^i$ moving in the physical spacetime. It is important to stress that $X^\mu(\sigma^a)$ is a dynamical variable and cannot be taken as the physical spacetime coordinate. The relation between $X^\mu(\sigma^a)$ and the physical spacetime coordinate will be made clear in later section.

Meanwhile, the line element \eqref{dl02} defining the background metric $g_{\mu\nu}(x)$ shall be promoted into the following one
\begin{align}
dl^2 & = g_{\mu\nu}(X) dX^\mu dX^\nu \nonumber \\
& = g_{\mu\nu}(X^\mu(\sigma)) \frac{\partial X^\mu}{\partial \sigma^a} \frac{\partial X^\nu}{\partial \sigma^b} d\sigma^a d \sigma^b \nonumber \\
& \equiv \mathcal G_{ab}(X^\mu(\sigma)) d\sigma^a d\sigma^b  \label{diffeo_transform_promotion}
\end{align}
where, by construction, the quantity $\mathcal G_{ab}(X^\mu(\sigma))$ is invariant under the following infinitesimal transformation
\begin{align}
X^\mu \to X^\mu - \xi^\mu(X), \qquad  g_{\mu\nu}(X) \to g_{\mu\nu}(X) + \nabla_\mu \xi_\nu + \nabla_\nu \xi_\mu. \label{diffeo_transform_enlarge}
\end{align}
There is a major difference between \eqref{diffeo_transform_enlarge} and \eqref{diffeo_trans}: the quantity $X^\mu$ is a dynamical variable while the $x^\mu$ is simply a label. With the symmetry \eqref{diffeo_transform_enlarge}, the dynamical equation of motion for $X^\mu$ is nothing but the conservation law of the stress tensor $T_{\mu\nu}$. The quantity $\mathcal G_{\mu\nu}(X)$ is the building block for formulating the local EFT for dissipative fluid. Indeed, the relation \eqref{diffeo_transform_promotion} will emerge naturally from the bulk analysis \cite{Crossley:2015tka} to be presented in next section.

Now, in the spirit of the Wilsonian RG, one can imagine integrating out all the modes but the hydrodynamic fields denoted by $X^\mu$, giving rise to a low energy EFT description of the original system. In other words, the original partition function \eqref{Z_micro} can be alternatively expressed as a path integral over the low energy hydrodynamic fields
\begin{align}
Z[g_1,\, g_2] = \int [D X_1^\mu] [D X_2^\nu] e^{i S_{eff}[\mathcal{G}_{1ab}[X], \,\, \mathcal{G}_{2ab}[X]]}, \label{Z_EFT}
\end{align}
where $S_{eff}$ is the hydrodynamic effective action. Here, we have recovered the SK indices for the external sources and the dynamical variables
\begin{align}
g_{\mu\nu} \to g_{1\mu\nu},\, g_{2\mu\nu}; \qquad X^\mu \to X_1^\mu, \, X_2^\mu; \qquad \mathcal G_{ab} \to \mathcal G_{1ab},\, \mathcal G_{2ab}.
\end{align}
Presumably, the local effective action $S_{eff}$ is of the form
\begin{align}
S_{eff} = \int d^4\sigma \sqrt{-\mathcal G_\r(\sigma)} \, \tilde{\mathcal L}_{eff}[\mathcal G_{\r ab}(\sigma),\, \mathcal G_{\a ab}(\sigma)] \label{Seff}
\end{align}
where we have introduced the Keldysh basis,
\begin{align}
\mathcal G_{\r ab} \equiv \frac{1}{2}(\mathcal G_{1ab} + \mathcal G_{2ab}), \qquad  \mathcal G_{\a ab} \equiv \mathcal G_{1ab} - \mathcal G_{2ab}
\end{align}
In \eqref{Seff}, $\mathcal G_\r(\sigma)$ denotes the determinant of the metric $\mathcal G_{\r ab}$. The volume element in \eqref{Seff} is motivated by the fact that $d^4\sigma \sqrt{-\mathcal G_\r(\sigma)} = d^4x \sqrt{-g_\r(x)}$ with $g_\r(x)$ the determinant of the external metric $g_{\r\mu\nu}(x)$. Notice that in \eqref{Seff} the local action is formulated in the fluid spacetime spanned by $\sigma^a$. Indeed, it is also possible to reformulate the EFT action in the physical spacetime, see \cite{Crossley:2015evo} for more details. We will elaborate on this point in later section.

Practically, it is very challenging (if not impossible) to derive $\tilde{\mathcal L}_{eff}[\mathcal G_{\r ab},\, \mathcal G_{\a ab}]$ by implementing the ``integrating out'' procedure for a generic quantum many-body system. However, with the hydrodynamic fields $X^\mu$ clearly identified, the hydrodynamic EFT can be formulated via proposing a set of symmetries \cite{Crossley:2015evo}, which we list below.

(1) Normalization condition
\begin{align}
S_{eff}[\mathcal G_{\r ab}, \mathcal G_{\a ab}=0] = 0. \label{normal_condition}
\end{align}

(2) $Z_2$ reflection symmetry
\begin{align}
S_{eff}[\mathcal G_{\r ab}, -\mathcal G_{\a ab}] = - \left( S_{eff}[\mathcal G_{\r ab}, \mathcal G_{\a ab}] \right)^*. \label{Z2_reflection}
\end{align}

(3) The imaginary part of $S_{eff}$ is non-negative
\begin{align}
{\rm Im}(S_{eff}) \geq 0. \label{PI_well-define}
\end{align}

(4) Re-parametrization symmetry in the fluid spacetime

The EFT action \eqref{Seff} is invariant under the following two independent transformations
\begin{align}
\sigma^i \to \sigma^{\prime i}(\sigma^i), ~~ \sigma^0 \to \sigma^0; \qquad {\rm or} \qquad \sigma^0 \to \sigma^{\prime 0}(\sigma^0, \sigma^i),~~ \sigma^i \to \sigma^i. \label{fluid_reparam}
\end{align}

(5) Dynamical Kubo-Martin-Schwinger (KMS) symmetry
\begin{align}
S_{eff}[\mathcal G_{\r ab}, \mathcal G_{\a ab}] = S_{eff}[\tilde{\mathcal G}_{\r ab}, \tilde{\mathcal G}_{\a ab}], \label{dynamical_KMS}
\end{align}
where
\begin{align}
\tilde{\mathcal G}_{\r ab}(-\sigma) = \mathcal G_{\r ab}(\sigma), \qquad  \tilde{\mathcal G}_{\a ab}(-\sigma) = \mathcal G_{\a ab}(\sigma) + {\rm i}\beta_0 \partial_0 \mathcal G_{\r ab}(\sigma),
\end{align}
where $\beta_0$ is the inverse temperature at spatial infinity.

The set of symmetries stringently constrains the form of EFT action. We briefly discuss them. The conditions \eqref{normal_condition} and \eqref{Z2_reflection} are the basic properties of the SK formalism. The condition \eqref{PI_well-define} ensures that the path integral \eqref{Z_EFT} is well-defined. The rest two symmetries \eqref{fluid_reparam} and \eqref{dynamical_KMS} will guide one to construct the effective action. The requirement \eqref{fluid_reparam} essentially defines what one means by a fluid \cite{Crossley:2015evo} (see also \cite{Dubovsky:2011sj}) and motivates to construct the building blocks from $\mathcal G_{\s ab}$ ($\s=1,2$), which transform as tensors with respect to the fluid re-parametrization symmetry \eqref{fluid_reparam}. This construction is rather technical and can be found in \cite{Crossley:2015evo}. The dynamical KMS symmetry \eqref{dynamical_KMS} incorporates statistical fluctuations, as required by the fluctuation-dissipation theorem. It reflects two facts on the physical system under consideration: it is in a thermal state and the microscopic theory is invariant under the time-reversal transformation. Ref.~\cite{Crossley:2015evo} realized that \eqref{dynamical_KMS} is indeed a classical statistical limit of a more general KMS symmetry at the quantum level.

In the holographic calculation to be presented in later sections, we will find it more convenient to split the building block $\mathcal G_{\s ab}$ as follows
\begin{align}
\mathcal G_{\s ab} = \eta_{ab} + B_{\s ab}, \qquad \s=1~{\rm or}~2. \label{bdy_data_split}
\end{align}
Then, the action \eqref{Seff} can be rewritten as
\begin{align}
S_{eff} = \int d^4\sigma \mathcal L_{eff}[B_{\r ab}, \, B_{\a ab}]
\end{align}
where $\mathcal L_{eff}$ will be derived through the holographic calculation. Here, $\mathcal L_{eff}$ will be presented in terms of $B_{\r ab}(\sigma)$ and $B_{\a ab}(\sigma)$ defined as
\begin{align}
B_{\r ab}(\sigma) \equiv \frac{1}{2}\left[ B_{1ab}(\sigma) + B_{2ab}(\sigma) \right], \qquad B_{\a ab}(\sigma) \equiv B_{1ab}(\sigma) - B_{2ab}(\sigma). \label{Bra_def}
\end{align}
Furthermore, we split the dynamical field and the external source as\footnote{With \eqref{X_g_linearization}, it is straightforward to obtain the perturbative expansion of the quantity $g_{\s \mu\nu}(X(\sigma))$.}
\begin{align}
X_\s^\mu (\sigma) = \delta_a^\mu \sigma^a + \pi_\s^\mu(\sigma), \qquad g_{\s\mu\nu}(x) = \eta_{\mu\nu} + A_{\s \mu\nu}(x). \label{X_g_linearization}
\end{align}
Viewing $\pi_{\s\mu}$ and $A_{\s \mu\nu}$ as perturbations of the same order, we can expand $B_{\s ab}$ of \eqref{bdy_data_split} as
\begin{align}
B_{\s ab}(\sigma) = & A_{\s ab}(\sigma) + A_{\s a\nu}(\sigma) \partial_b \pi_\s^\nu(\sigma) + A_{\s\mu b}(\sigma) \partial_a \pi_\s^\mu(\sigma) + \pi_\s^\alpha(\sigma) \partial_\alpha A_{\s ab}(\sigma) \nonumber \\
& + \partial_a \pi_{\s b}(\sigma) + \partial_b \pi_{\s a}(\sigma) + \partial_a \pi_\s^\mu(\sigma) \partial_b \pi_{\s\mu}(\sigma) + \cdots  \label{B12_expand}
\end{align}
In \eqref{B12_expand}, we have truncated the expansion at quadratic order, which will be found sufficient for capturing the Gaussian terms in boundary action. Then, \eqref{Bra_def} are expanded as
\begin{align}
B_{\r ab}(\sigma)& = A_{\r ab}(\sigma) + \partial_a \pi_{\r b}(\sigma) + \partial_b \pi_{\r a}(\sigma) + A_{\r a\nu}(\sigma) \partial_b \pi_\r^\nu(\sigma) + \frac{1}{2} A_{\a a\nu}(\sigma) \partial_b \pi_\a^\nu(\sigma) \nonumber \\
& + A_{\r \mu b}(\sigma) \partial_a \pi_\r^\mu(\sigma) + \frac{1}{2} A_{\a \mu b}(\sigma) \partial_a \pi_\a^\mu(\sigma) + \pi_\r^\alpha(\sigma) \partial_\alpha A_{\r ab}(\sigma) + \frac{1}{2} \pi_\a^\alpha(\sigma) \partial_\alpha A_{\a ab}(\sigma) \nonumber \\
& + \partial_a \pi_\r^\mu(\sigma) \partial_b \pi_{\r \mu}(\sigma) + \frac{1}{2} \partial_a \pi_\a^\mu(\sigma) \partial_b \pi_{\a\mu}(\sigma)+ \cdots, \nonumber \\
B_{\a ab}(\sigma)& = A_{\a ab}(\sigma) + \partial_a \pi_{\a b}(\sigma) + \partial_b \pi_{\a a}(\sigma) + A_{\r a\nu}(\sigma) \partial_b \pi_\a^\nu(\sigma) + A_{\a a\nu}(\sigma) \partial_b \pi_\r^\nu(\sigma) \nonumber \\
& + A_{\r\mu b}(\sigma) \partial_a \pi_\a^\mu(\sigma) + A_{\a\mu b}(\sigma) \partial_a \pi_\r^\mu(\sigma) + \pi_\r^\alpha(\sigma) \partial_\alpha A_{\a ab}(\sigma) + \pi_\a^\alpha(\sigma) \partial_\alpha A_{\r ab}(\sigma) \nonumber \\
& + \partial_a \pi_\r^\mu(\sigma) \partial_b \pi_{\a\mu}(\sigma) + \partial_a \pi_\a^\mu(\sigma) \partial_b \pi_{\r\mu}(\sigma) + \cdots, \label{Bra_expand}
\end{align}
which will be useful in rewriting the holographic action in the physical spacetime.

\subsection{The holographic perspective} \label{holo_RG}

The holographic duality makes it possible to derive the effective action for a dissipative fluid whose underlying microscopic dynamics involves a strongly coupled field theory. Basically, this amounts to a holographic Wilsonian RG involving the classical dynamics of the AdS gravity. In this section, we outline such a program. The discussion will closely follow \cite{Crossley:2015tka}.

The starting point is the holographic dictionary:
\begin{align}
Z_{\rm CFT} = Z_{\rm AdS}
\end{align}
Here, once identified as $Z$ of \eqref{Z_EFT}, the CFT partition function $Z_{\rm CFT}$ is expressed as a path integral over the low energy variable $X$ (both the Lorentzian and SK indices are suppressed)
\begin{align}
Z_{\rm CFT} = \int [DX] e^{i S_{eff}[X]} \label{Z_CFT}
\end{align}
where $S_{eff}[X]$ is the hydrodynamic effective action that we are looking for. The AdS partition function $Z_{\rm AdS}$ involves a path integral over the bulk metric field
\begin{align}
Z_{\rm AdS} = \int [DG] e^{iS[G]}, \label{Z_AdS}
\end{align}
where $S[G]$ is the total bulk action. Then, applying the idea of the Wilsonian RG to \eqref{Z_AdS}, one could integrate out the heavy modes (dual to the gapped modes of the boundary CFT) while leave aside the light modes (corresponding to the hydrodynamic variable $X$). This holographic analogue of the Wilsonian RG is supposed to bring the AdS partition function \eqref{Z_AdS} into the desired form \eqref{Z_CFT}. Thus, the derivation of the hydrodynamic EFT from the AdS gravity boils down to identifying the gravity dual of the gapless modes and integrating out the heavy modes through the bulk calculation.

In the holographic context, the identification of the hydrodynamic variables was nicely clarified by Nickel and Son \cite{Nickel:2010pr}, and then further refined in \cite{Crossley:2015tka} (see also \cite{deBoer:2015ija}). Analogous to the field theory discussion as summarized in section \ref{idea_field_theory}, the basic idea is to play with the diffeomorphism symmetry of the bulk gravity \cite{Crossley:2015tka}.

First, we write down a general bulk metric
\begin{align}
ds^2 = \tilde G_{AB}(x) dx^{A} dx^{B} \equiv N^2 dz^2 + \chi_{\mu\nu} (dx^\mu + N^\mu dz) (dx^\nu + N^\nu dz) \label{metric_no_gauge-fixing}
\end{align}
where $x^A = (z, x^\mu)$. The UV (AdS boundary) and IR (horizon) hypersurfaces are some constant-$z$ slices:
\begin{align}
\Sigma_{\rm UV}: \quad z= z_{\rm UV}; \qquad \qquad \Sigma_{\rm IR}: \quad z=z_{\rm IR}.
\end{align}
Then, up to a conformal factor, $\chi_{\mu\nu}$ at the AdS boundary is nothing but the external metric for the boundary system
\begin{align}
\chi_{\mu\nu}(z=z_{\rm UV}, x^\mu) \propto g_{\mu\nu}(x^\mu).
\end{align}

We know that the metric \eqref{metric_no_gauge-fixing} contains ``unphysical'' gauge degrees of freedom. For instance, thanks to the bulk diffeomorphism invariance, it is always possible to remove $N$ and $N^\mu$, while treat $\chi_{\mu\nu}$ as the dynamical degrees of freedom in the bulk. To advance, we consider a bulk coordinate transformation bringing \eqref{metric_no_gauge-fixing} to the following form,
\begin{align}
ds^2 = \hat G_{MN}(y) dy^M dy^N = \hat G_{rr}[\hat G_{\mu\nu}] dr^2 + 2 \hat G_{r\mu}[\hat G_{\mu\nu}] dr dy^\mu + \hat G_{\mu\nu} dy^\mu dy^\nu, \label{metric_gauge_fixed}
\end{align}
where $y^M = (r, y^\mu)$. Here, the gauge condition has been specified implicitly such that $\hat G_{rM}$ are completely fixed in terms of $\hat G_{\mu\nu}$. Therefore, only $\hat G_{\mu\nu}$ are dynamical fields in the bulk. In \cite{Crossley:2015tka}, a specific gauge condition is taken as $\hat G_{rr}=1, \hat G_{r\mu}=0$.

Going from $x^A$ to $y^M$, we have the change rule for the bulk metric:
\begin{align}
\hat G_{MN}(y) = \tilde G_{AB}(x) \frac{\partial x^A} {\partial y^M} \frac{\partial x^B} {\partial y^N}. \label{gauge_transform}
\end{align}
Then, the implicit gauge condition (i.e., fixing $\hat G_{rM}$) reads
\begin{align}
\tilde G_{AB}(x) \frac{\partial x^A} {\partial r} \frac{\partial x^B}{\partial r} = \hat G_{rr}(x), \nonumber \\
\tilde G_{AB}(x) \frac{\partial x^A} {\partial r} \frac{\partial x^B}{\partial y^\mu} = \hat G_{r\mu}(y)  \label{gauge_eom}
\end{align}
Apparently, \eqref{gauge_eom} could be understood as the differential equations for the functions $r(x^A)$ and $y^\mu(x^A)$. Indeed, if we follow \cite{Crossley:2015tka} and choose the gauge condition $\hat G_{rr}=1, \hat G_{r\mu}=0$, this statement will become more obvious.

Moreover, just as in solving the gauge transformation parameter in the $U(1)$ example, suitable boundary conditions at $\Sigma_{\rm UV}$ and/or $\Sigma_{\rm IR}$ are needed to fully determine the ``gauge transformation parameters'' $r(x^A)$ and $y^\mu(x^A)$. Practically, we will follow \cite{Crossley:2015tka} and take a hybrid fixing\footnote{More generally, we could have taken $r_0 = r_0(x^\alpha)$ as for $y^\mu|_{\Sigma_{\rm IR}}$. However, the diffeomorphism symmetry on the AdS boundary allows us to render $r_0$ to be a constant.}:
\begin{align}
r|_{\Sigma_{\rm UV}} = r_0 = {\rm constant}, \qquad y^\mu|_{\Sigma_{\rm IR}} = \sigma^a(x^\alpha) \delta_a^\mu, \label{hybrid_fixing}
\end{align}
where the constant $r_0$ will be actually taken as $+\infty$. The values of $r(x^A)$ and $y^\mu(x^A)$ at the other ends are determined {\it dynamically}, via solving \eqref{gauge_transform}. In general, the results can be parameterized as (the $r_0$-dependence is made implicit)
\begin{align}
r|_{\Sigma_{\rm IR}} = \tilde \tau(\sigma^a), \qquad y^\mu|_{\Sigma_{\rm UV}} = X^\mu(\sigma^a). \label{r_ymu_other_ends}
\end{align}
Clearly, in the coordinate system \eqref{metric_gauge_fixed}, the UV and IR hypersurfaces are specified as (in terms of the values for $r$)
\begin{align}
\Sigma_{\rm UV}: \quad r=r_0; \qquad  \Sigma_{\rm IR}: \quad r = \tilde \tau (X^{-1}(y^\mu)) \equiv \tau(y^\mu)
\end{align}
Interestingly, from \eqref{hybrid_fixing} and \eqref{r_ymu_other_ends}, we read off the relative embedding between $\Sigma_{\rm UV}$ and $\Sigma_{\rm IR}$, described as $X^\mu(\sigma^a)$, and the proper distance between $\Sigma_{\rm UV}$ and $\Sigma_{\rm IR}$ (i.e., the $\tilde \tau(\sigma^a)$). In \eqref{hybrid_fixing}, we see the emergence of the fluid spacetime spanned by $\sigma^a$.

In analog with the U(1) case \cite{Nickel:2010pr,Crossley:2015tka,Bu:2020jfo}, we can imagine that the solutions for $r$ and $y^\mu$ can be written as the gauge links of $N$ and $N^\mu$. Implicitly, the dynamical field $X^\mu(\sigma)$ is like the Wilson line of $N^\mu$ along a certain path. This conclusion was actually demonstrated more transparently in the linearized bulk theory, see \cite{deBoer:2015ija} for more details.

From \eqref{gauge_transform}, we can read off the relation between intrinsic metrics on the UV hypersurface\footnote{In the first line of \eqref{intrinsic_metric}, the $g$ on the left-handed side is easily understood by taking the limit $X^\mu \to x^\mu$.},
\begin{align}
g_{\mu\nu}(X^\mu) & = g_{\alpha\beta}(x) \frac{\partial x^\alpha}{\partial X^\mu} \frac{\partial x^\beta}{\partial X^\nu} \nonumber \\
& = g_{\alpha\beta}(x) \frac{\partial x^\alpha}{\partial \sigma^a} \frac{\partial \sigma^a}{\partial X^\mu} \frac{\partial x^\beta}{\partial \sigma^b}  \frac{\partial \sigma^b}{\partial X^\nu} \nonumber \\
& = \mathcal G_{ab}(\sigma) \frac{\partial \sigma^a}{\partial X^\mu} \frac{\partial \sigma^b}{\partial X^\nu} \label{intrinsic_metric}
\end{align}
where in the second line we have made use of the chain rule given that $X^\mu = X^\mu(\sigma^a)$ and $\sigma^a = \sigma^a(x^\alpha)$; in the last line we considered the coordinate transformation $x^\mu \to \sigma^a$, $g_{\alpha\beta}(x) \to \mathcal G_{ab}(\sigma)$. Then, with the chain rule, we have
\begin{align}
ds^2|_{\Sigma_{\rm UV}} \simeq dl^2 & = g_{\mu\nu}(X) dX^\mu dX^\nu = g_{\mu\nu}(X^\mu(\sigma)) \frac{\partial X^\mu}{\partial \sigma^a} \frac{\partial X^\nu}{\partial \sigma^b} d\sigma^a d\sigma^b \nonumber \\
& = \mathcal G_{ab}(\sigma) d\sigma^a d\sigma^b \label{bdy_data}
\end{align}
which is obviously identical to \eqref{diffeo_transform_promotion}. When solving the bulk dynamics, we will take \eqref{bdy_data} as the boundary condition for the bulk metric. It it important to stress that, with certain gauge-fixing assumed in \eqref{metric_gauge_fixed}, the AdS boundary condition for the bulk metric encodes not only the external metric but also the hydrodynamic field on the boundary.

Thus, solving \eqref{gauge_eom} under the boundary conditions \eqref{hybrid_fixing}, we are supposed to get $y^\mu = y^\mu(\sigma^a)$. Plugging this solution into \eqref{metric_gauge_fixed}, we are motivated to rewrite the line element \eqref{metric_gauge_fixed} in the emergent spacetime spanned by $\sigma^M = (r, \sigma^a)$
\begin{align}
ds^2 = G_{MN}(\sigma) d\sigma^M d\sigma^N = G_{rr}[G_{ab}] dr^2 + 2 G_{ra}[G_{ab}] dr d\sigma^a + G_{ab} d\sigma^a d\sigma^b, \label{sigma_coord}
\end{align}
where, at the AdS boundary, $G_{ab}$ will be $\mathcal G_{ab}(\sigma)$ of \eqref{bdy_data} up to a conformal factor.

Before proceeding, we briefly summarize the three sets of coordinate systems that we have introduced. The first one is $(x^A, \tilde G_{AB})$ in which we did not make any gauge-fixing. This coordinate system corresponds to the physical spacetime for the boundary theory, in which we clearly defined the external metric $g_{\mu\nu}(x)$. The second one is denoted by $(y^M, \hat G_{MN})$, which assumes certain gauge-fixing. Intriguingly, the coordinates $y^M = (r, y^\mu)$ are the dynamical variables. This corresponds to the $X^\mu$-coordinate system of the boundary analysis. While this coordinate system is intuitive in understanding the identification of the hydrodynamical field and emergence of the fluid spacetime, it is not convenient for practical calculations. Therefore, we further introduced a third coordinate system denoted by $\sigma^M = (r, \sigma^a)$, where $\sigma^a$ exactly spans the fluid spacetime. This third one is friendly for the bulk calculations. Instead of treating the coordinate as a dynamical field as in the second set of coordinate system, the dynamical variable is naturally encoded in the AdS boundary condition.

The discussion above implies two equivalent ways of counting the bulk degrees of freedom (both the gauge and dynamical ones)
\begin{align}
\tilde G_{AB}=(N, N^\mu, \chi_{\mu\nu}) \Leftrightarrow (r, y^\mu, \hat G_{\mu\nu})
\end{align}
Therefore, we have two equivalent ways of expressing the AdS partition function \eqref{Z_AdS}: a path integral over the bulk metrics $(N, N^\mu, \chi_{\mu\nu})$ versus a path integral over $(r, y^\mu, \hat G_{\mu\nu})$:
\begin{align}
Z_{\rm AdS} &= \int [D N][D N^\mu][D\chi_{\mu\nu}] e^{iS[N, N^\mu, \chi_{\mu\nu}]} \nonumber \\
&= \int [D r][D y^\mu][D \hat G_{\mu\nu}] e^{iS[\hat G_{\mu\nu}]} \nonumber \\
&\simeq \int [D \tau] [D X^\mu] [D G_{ab}] e^{i S[G_{ab}[\sigma^M; \tau, X^\mu]]}  \nonumber \\
& = \int [D \tau][D X^\mu]e^{iS[G_{ab}[X^\mu]]|_{\rm p.o.s}} \nonumber \\
& = \int [D X^\mu] e^{iS[G_{ab}[X^\mu]]|_{\rm p.o.s}}. \label{Z_AdS1}
\end{align}
We elaborate on \eqref{Z_AdS1} step by step. In the first line, the bulk action is an integral in $x^M$-coordinate. The second line seems odd: on the one hand, the bulk action is defined as an integral over the coordinates $(r,y^\mu)$; on the other hand, the coordinates $(r,y^\mu)$ are the dynamical variables which are integrated over in the path integral. Indeed, via the transformation from \eqref{metric_gauge_fixed} to \eqref{sigma_coord}, the second line shall be understood in terms of the third line, where the integration over $(r, y^\mu)$ was converted to an integration over the hydrodynamic variables $X^\mu$ as well as $\tau$. In addition, the bulk action is an integral in the $\sigma$-coordinate. Then, from the third line to the fourth line, $G_{ab}$ (the gravity dual of the heavy modes) was integrated out in the saddle point approximation. This is simply achieved via substituting the classical solution for $G_{ab}$ into the bulk action $S[G_{ab}]$. Here, the subscript ``$_{\rm p.o.s}$'' stands for partially on-shell, which emphasizes that only the dynamical equations will be solved to obtain the solution for $G_{ab}$. This partially on-shell prescription can be consolidated by revising the gravitational variational problem based on gauge-fixed configuration \eqref{metric_gauge_fixed} or \eqref{sigma_coord}, see section \ref{gravity_var_problem} for more details. Note that the bulk action $S[G_{ab}[X^\mu]]$ depends on $X^\mu$ through the boundary data $\mathcal G_{ab}(\sigma)$, which implies that $X$ is the gapless mode (it enters via the derivative $\partial_a X^\mu$). Moreover, the partially on-shell action would not depend on the variable $\tau$, which indicates the integration over $\tau$ will give an infinite constant and could be dropped without affecting physical observable. Thus, the AdS partition function is eventually cast into the desired form of \eqref{Z_CFT} with
\begin{align}
S_{eff} = S[G_{ab}[X^\mu]]|_{\rm p.o.s}
\end{align}

Finally, the analysis above shall be extended to the situation of double copy AdS so that boundary theory lives on the SK closed time path, which is in parallel with the field theory consideration in section \ref{idea_field_theory} and shall go smoothly.

\subsection{Variational problem of gravity revisited} \label{gravity_var_problem}

In this section we revisit the variational problem of gravity. Usually, the variational problem is done without making a gauge-fixing. However, it was realized that \cite{Crossley:2015tka} variational problem can be made well-posed using a gauge-fixed configuration as long as the gauge degrees of freedom are carefully treated. Interestingly, the latter approach will demonstrate validity of the partially on-shell prescription for solving the bulk dynamics, as alluded to below \eqref{Z_AdS1}.

Recall that our notation convention is as follows: $(x^A, \tilde G_{AB})$ is for a coordinate system without any gauge-fixing, see \eqref{metric_no_gauge-fixing}; $(\sigma^M, G_{MN})$ is reserved for a coordinate system assuming certain gauge condition, see \eqref{sigma_coord}.

First, we consider the bulk variational problem in terms of $\tilde G_{AB}$, which can be varied freely. Then, we have
\begin{align}
\delta S = \frac{1}{2} \int d^5x \sqrt{-\tilde G} \tilde E^{AB} \delta \tilde G_{AB} \label{deltaS}
\end{align}
Here, a potential boundary term arising from $\delta S_0$ is exactly cancelled by $\delta S_{\rm GH}$. Thus, we obtain the bulk equations of motion (EOMs) for $\tilde G_{AB}$
\begin{align}
\delta \tilde G_{AB} \neq 0 \Rightarrow \tilde E^{AB} = 0
\end{align}
When $\delta \tilde G_{AB}$ is a coordinate transformation, i.e., $\delta \tilde G_{AB} = \tilde \nabla_A \delta x_B + \tilde \nabla_B \delta x_A$, we immediately obtain the Bianchi identities
\begin{align}
\delta x_B \neq 0 \Rightarrow \tilde \nabla_A \tilde E^{AB} = 0
\end{align}

Now, we turn to the bulk variational problem in terms of the field configuration $G_{MN}$. Note that in order to not miss the gauge degrees of freedom, we shall also take into account the variation of the bulk coordinate $\sigma^M$. Then, from \eqref{gauge_transform} (with the replacement $y\to \sigma$) we have
\begin{align}
\delta \tilde G_{AB} & = \delta G_{MN} \frac{\partial \sigma^M}{\partial x^A} \frac{\partial \sigma^N}{\partial x^B} + \frac{\partial G_{MN}}{\partial \sigma^P} \frac{\partial \sigma^M}{\partial x^A} \frac{\partial \sigma^N}{\partial x^B}\, \delta \sigma^P + 2 G_{MN} \frac{\partial \delta \sigma^M}{\partial x^A} \frac{\partial \sigma^N}{\partial x^B} \nonumber \\
& = \delta G_{MN} \frac{\partial \sigma^M}{\partial x^A} \frac{\partial \sigma^N} {\partial x^B} + \frac{\partial G_{MN}}{\partial \sigma^P} \frac{\partial \sigma^M} {\partial x^A} \frac{\partial \sigma^N}{\partial x^B}\, \delta \sigma^P + 2 G_{MN} \frac{\partial \sigma^P}{\partial x^A} \frac{\partial \sigma^N}{\partial x^B} \frac{\partial \delta \sigma^M}{\partial \sigma^P} \label{deltaGAB_deltaGMN}
\end{align}
However, as seen from \eqref{sigma_coord}, while the metric components $G_{ab}$ can be freely varied, the variation of $G_{rM}$ is completely fixed in terms of $\delta G_{ab}$. In practice, we will choose
\begin{align}
\delta G_{rr} = \frac{\delta G_{00}}{r^4f^2(r)}, \qquad \delta G_{r0} = - \frac{\delta G_{00}}{r^2 f(r)}, \qquad \delta G_{ri} = -  \frac{\delta G_{0i}}{r^2 f(r)}, \label{gauge_fixing_deltaGMN}
\end{align}
where $f(r)$ will be specified later, see the context below \eqref{metric_HMN}. Indeed, the condition \eqref{gauge_fixing_deltaGMN} follows from our gauge condition \eqref{gauge_condition_HMN}. With \eqref{deltaGAB_deltaGMN} and \eqref{gauge_fixing_deltaGMN}, the action variation \eqref{deltaS} is split into two parts
\begin{align}
\delta S = \delta S_1 + \delta S_2
\end{align}

The first part $\delta S_1$ arises from the metric variation $\delta G_{ab}$
\begin{align}
\delta S_1 & = \frac{1}{2} \int d^5\sigma \sqrt{-G} \tilde E^{AB} \frac{\partial \sigma^M}{\partial x^A} \frac{\partial \sigma^N}{\partial x^B} \delta G_{MN} \nonumber \\
& = \frac{1}{2} \int d^5\sigma \sqrt{-G} E^{MN} \delta G_{MN} \nonumber \\
& = \frac{1}{2} \int d^5\sigma \sqrt{-G} \left\{ \left[ E^{00} + \frac{E^{rr}}{r^4f^2(r)} - 2\frac{E^{r0}}{r^2 f(r)} \right] \delta G_{00} \right. \nonumber \\
& \qquad \qquad \qquad \qquad \left.+ 2 \left[ E^{0i} - \frac{E^{ri}}{r^2f(r)} \right] \delta G_{0i} + E^{ij} \delta G_{ij} \right\},
\end{align}
where the gauge choice of \eqref{gauge_fixing_deltaGMN} was substituted. Thus, we obtain
\begin{align}
& \delta G_{00} \neq 0 \Rightarrow E^{00} + \frac{E^{rr}}{r^4f^2(r)} - 2\frac{E^{r0}}{r^2 f(r)}=0, \nonumber \\
& \delta G_{0i} \neq 0 \Rightarrow E^{0i} - \frac{E^{ri}}{r^2f(r)} =0, \nonumber \\
& \delta G_{ij} \neq 0 \Rightarrow E^{ij} =0, \label{dynamical_EOMs}
\end{align}
which are the dynamical EOMs for the bulk gravity.

The second part $\delta S_2$ is generated by the variation of the coordinate $\delta \sigma^M$ and contains both the bulk and boundary terms
\begin{align}
\delta S & = \frac{1}{2} \int d^5\sigma \sqrt{-G} E^{MN} \left[\frac{\partial G_{MN}} {\partial \sigma^P} \delta \sigma^P + 2 G_{MP} \frac{\partial \delta \sigma^P}{\partial \sigma^N} \right] \nonumber \\
& = \frac{1}{2} \int d^5 \sigma \sqrt{-G} \left[ \nabla_N \left( E^{MN} 2 G_{MP} \delta \sigma^P \right) - \nabla_N\left( E^{MN} 2 G_{MP} \right)\delta \sigma^P \right] \nonumber \\
& = - \int d^5 \sigma \sqrt{- G}\, \nabla_N E^N_M \,\delta \sigma^M + \int d^4\sigma \sqrt{- \gamma} \, n_N E^N_M \, \delta \sigma^M \big|_{\rm boundary}. \label{deltaS_deltaxM}
\end{align}
The bulk part of \eqref{deltaS_deltaxM} gives the Bianchi identities
\begin{align}
\delta \sigma^M \neq 0 \Rightarrow  \nabla_N E^N_M = 0. \label{Bianchi_ID}
\end{align}
The boundary part of \eqref{deltaS_deltaxM} gives the constraint equations
\begin{align}
\delta \sigma^M \big|_{\rm boundary} \neq 0 \Rightarrow \sqrt{- \gamma} \, n_N E^N_M \big|_{\rm boundary} = 0 \label{constraint}
\end{align}
which contain both the Hamiltonian constraint ($M=r$) and the momentum constraints ($M=\mu$). Physically, the momentum constraints ($M=\mu$) of \eqref{constraint} correspond to the conservation law of the energy-momentum tensor on the boundary (which can be cast into the dynamical EOMs for $X^\mu$ in the hydrodynamic EFT), while the Hamiltonian constraint ($M=r$) of \eqref{constraint} will be satisfied automatically once the dynamical EOMs \eqref{dynamical_EOMs} and the momentum constraints are imposed. The latter fact is due to the Bianchi identities \eqref{Bianchi_ID}.

Therefore, we see emergence of the partially on-shell prescription for evaluating the AdS partition function as outlined below \eqref{Z_AdS1}. Solving the dynamical EOMs \eqref{dynamical_EOMs}, we will fully determine the bulk metric $G_{ab}$. In the saddle point approximation, we can integrate out the gapped modes $G_{ab}$ by plugging its classical solution into the total bulk action $S[G_{ab}]$, giving rise to the AdS partition function in the last line of \eqref{Z_AdS1}.

\section{Holographic derivation of hydrodynamic effective action} \label{holo_EFT}

In this section we set up the bulk perturbation theory and derive the boundary effective action by solving the classical dynamics of the bulk gravity.

\subsection{Perturbation theory in the bulk} \label{perturb_theory}

We consider a pure gravity with the total bulk action
\begin{align}
S= S_0 + S_{\rm GH} + S_{\rm ct}. \label{S}
\end{align}
The bulk term $S_0$ is the standard Einstein-Hilbert action plus a negative cosmological constant:
\begin{align}
S_0 = \frac{1}{16\pi G_5} \int d^5\sigma \sqrt{-G} \left( R[G] - 2 \Lambda \right), \label{S0}
\end{align}
where $\Lambda = - 6/L^2$ with $L$ the AdS curvature radius. Hereafter, $L$ will be set to unity for simplicity. The variation of the bulk term $S_0$ gives the Einstein equations
\begin{align}
E^{MN} \equiv R^{MN}[G] - \frac{1}{2} G^{MN} R[G] - 6 G^{MN} =0. \label{EMN}
\end{align}

In order to ensure the variational problem of the gravity to be well-defined, see \eqref{deltaS}, we need the boundary term $S_{\rm GH}$
\begin{align}
S_{\rm GH} = \frac{1}{8\pi G_5}\int_{\Sigma_{\rm UV}} d^4\sigma \sqrt{-\gamma} K[\gamma] \label{S_GH}
\end{align}
where $\gamma$ is the determinant of the induced metric $\gamma_{ab}$ on the boundary hypersurface $\Sigma_{\rm UV}$. $K[\gamma]$ is the extrinsic scalar curvature of the induced metric $\gamma_{ab}$.

The counter-term $S_{\rm ct}$ is \cite{deHaro:2000vlm,Balasubramanian:1999re}
\begin{align}
S_{\rm ct} = -\frac{1}{16\pi G_5} \int_{\Sigma_{\rm UV}} d^4\sigma \sqrt{-\gamma} \left\{ 6 + \frac{1}{2} R[\gamma] - \left(\frac{1}{8} R^{ab}[\gamma]R_{ab}[\gamma] - \frac{1}{24} R^2[\gamma] \right) \log{\frac{1}{r^2}}\right\} \label{Sct}
\end{align}
where we assumed a minimal subtraction scheme. The logarithmic term arises from the conformal anomaly and is necessary when the boundary system is put in a curved spacetime. Hereafter, we will set $8\pi G_5 = 1$ for convenience.

Note that both $S_{\rm GH}$ and $S_{\rm ct}$ are defined near the boundary hypersurface $\Sigma_{\rm UV}$ with $r\to \infty$ assumed implicitly. Since $r$ is bounded by two asymptotic boundaries, we have one copy of $S_{\rm GH}$ and $S_{\rm ct}$ at each boundary hypersurface with $r=\infty_\s$ ($\s=1~{\rm or}~2$).

We split the bulk metric \eqref{sigma_coord} into two parts
\begin{align}
ds^2 & = \bar G_{MN} d\sigma^M d\sigma^N + H_{MN} d\sigma^M d\sigma^N \nonumber \\
& = 2drd\sigma^0 - r^2 f(r) (d\sigma^0)^2 + r^2 \delta_{ij} d\sigma^i d\sigma^j + H_{MN} d\sigma^M d\sigma^N \label{metric_HMN}
\end{align}
where $f(r) = 1-r_h^4/r^4$. For $H_{MN}$ we assume the following gauge condition
\begin{align}
H_{rr} = \frac{H_{00}}{r^4f^2(r)}, \qquad  H_{r0} = - \frac{H_{00}}{r^2f(r)}, \qquad  H_{ri} = - \frac{H_{0i}}{r^2f(r)},  \label{gauge_condition_HMN}
\end{align}
which immediately implies the condition \eqref{gauge_fixing_deltaGMN}. Here, $\bar G_{MN}$ in \eqref{metric_HMN} represents the Schwarzschild-AdS$_5$ black brane, which corresponds to a finite temperature state for the boundary system. In order to put the boundary system on the SK closed time path of Figure \ref{SK contour}, the radial coordinate $r$ varies along the contour of Figure \ref{holographic_SK_contour}. The bulk metric perturbation $H_{MN}$ is dual to the fluctuations and dissipations of the boundary system. In order to solve the bulk perturbation, we will impose Dirichlet-type boundary condition
\begin{align}
H_{ab}(r\to \infty_\s, \sigma^a) = r^2 \left[ B_{\s ab}(\sigma) + \mathcal{O}(r^{-1}) \right], \label{Hab_bdy_condition}
\end{align}
where $B_{\s ab}$ is introduced in \eqref{bdy_data_split}. In the linearization approximation, the perturbative expansion of $B_{\s ab}$ has been presented in \eqref{B12_expand}. Actually, with \eqref{Hab_bdy_condition}, we have the desired AdS condition for the total bulk metric
\begin{align}
G_{ab}(r\to \infty_\s, \sigma^a) = r^2 \left[ \mathcal G_{\s ab}(\sigma) + \mathcal{O}(r^{-1}) \right]
\end{align}
with the boundary data $\mathcal G_{\s ab}(\sigma)$ split as in \eqref{bdy_data_split}.

Given the highly nonlinear feature of the Einstein gravity, we consider expanding the bulk action $S_0$ of \eqref{S0} in the number of the bulk perturbation $H_{MN}$
\begin{align}
S_0 & = S_0^{(0)} + S_0^{(1)} + S_0^{(2)} + S_0^{(3)} + \cdots \nonumber \\
& = \int d^5\sigma \sqrt{- \bar G} \left[ {\mathcal L}_0^{(0)} + \mathcal L_0^{(1)} + \mathcal L_0^{(2)} + \mathcal L_0^{(3)} + \cdots \right], \label{S0_expansion}
\end{align}
where a superscript denotes the number of $H_{MN}$. The various parts are
\begin{align}
{\mathcal L}_0^{(0)} = - 4, \label{L0_(0)}
\end{align}
\begin{align}
\mathcal L_0^{(1)} =  \frac{1}{2} \bar \nabla_M \bar \nabla_N H^{MN} - \frac{1}{2} \bar \nabla^2 H, \label{L0_(1)}
\end{align}
\begin{align}
\mathcal L_0^{(2)} = & -2 H_{MN} H^{MN} + H^2 + H^{MN} \bar \nabla_M \bar \nabla_N H - H^{MN} \bar \nabla_N \bar \nabla_A H_M^A \nonumber \\
&- \frac{1}{4} \bar \nabla_M H \bar \nabla^M H - \bar \nabla_M H^{MN} \bar \nabla_A H_N^A + \bar \nabla^M H \bar \nabla_N H_M^N \nonumber \\
& - H^{MN} \bar \nabla_A \bar \nabla_N H_M^A + \frac{1}{2} H \bar \nabla_M \bar \nabla_N H^{MN} + H^{MN} \bar \nabla^2 H_{MN} \nonumber \\
& - \frac{1}{2} H \bar \nabla^2 H - \frac{1}{2} \bar \nabla_M H_{N A} \bar \nabla^A H^{MN} + \frac{3}{4} \bar \nabla_A H_{MN} \bar \nabla^A H^{MN}, \label{L0_(2)}
\end{align}
\begin{align}
\mathcal L_{0}^{(3)} = &\frac{8}{3} H_{M}^{B} H^{MN} H_{NB} + \frac{1}{3} H^3 -2 H H_{MN} H^{MN} -\frac{3}{4} H^{MN} \bar \nabla_{M} H^{AB} \bar \nabla_{N} H_{AB} \nonumber \\
& + \frac{1}{4} H^{MN} \bar \nabla_{M} H \bar \nabla_{N} H - H^{MN} \bar \nabla_{A} H_{M}^{A} \bar \nabla_{N} H -H^{MN} \bar \nabla_{N} H_{M}^{A} \bar \nabla_{A} H \nonumber \\
& + \frac{1}{2} H^{MN} \bar \nabla_{A} H_{MN} \bar \nabla^{A} H -\frac{1}{8} H \bar \nabla_{C} H \bar \nabla^{C} H +H^{MN} \bar \nabla_{A} H_{M}^{A} \bar \nabla_{B} H_{N}^{B} \nonumber \\
& +2 H^{MN} \bar \nabla_{N} H_{M}^{A} \bar \nabla_{B} H_{A}^{B} - \frac{1}{2} H \bar \nabla_{N} H^{NA} \bar \nabla_{B} H_{A}^{B} -H^{MN} \bar \nabla_{A} H_{MN} \bar \nabla_{B} H^{AB} \nonumber \\
& +\frac{1}{2} H \bar \nabla_{A} H \bar \nabla_{B} H^{AB} + H^{MN} \bar \nabla_{N} H_{AB} \bar \nabla^{B} H_{M}^{A} +\frac{1}{2} H^{MN} \bar \nabla_{A} H_{BN} \bar \nabla^{B} H_{M}^{A}  \nonumber \\
&-\frac{3}{2} H^{MN} \bar \nabla_{A} H_{BN} \bar \nabla^{A} H_{M}^{B} -\frac{1}{4} H \bar \nabla_{A} H_{BN} \bar \nabla^{B} H^{AN} +\frac{3}{8} H \bar \nabla_{A} H_{BN} \bar \nabla^{A} H^{BN} \nonumber \\
&-H_{M}^{A} H^{MN} \bar \nabla_{A} \bar \nabla_{N} H + \frac{1}{2} H H^{AN} \bar \nabla_{A} \bar \nabla_{N} H + H_{M}^{A} H^{MN} \bar \nabla_{A} \bar \nabla_{B} H_{N}^{B} \nonumber \\
&-\frac{1}{2} H H^{AN} \bar \nabla_{A} \bar \nabla_{B} H_{N}^{B} + H^{MN} H^{AB} \bar \nabla_{B} \bar \nabla_{N} H_{AM} - H^{MN} H^{AB} \bar \nabla_{B} \bar \nabla_{A} H_{MN}  \nonumber \\
&+H_{M}^{A} H^{MN} \bar \nabla_{B} \bar \nabla_{A} H_{N}^{B} -\frac{1}{2} H H^{AN} \bar \nabla_{B} \bar \nabla_{A} H_{N}^{B} -\frac{1}{4} H^{MN} H_{MN} \bar \nabla_{B} \bar \nabla_{A} H^{AB} \nonumber \\
&+ \frac{1}{8} H^2 \bar \nabla_{B} \bar \nabla_{A} H^{AB} - H_{M}^{A} H^{MN} \bar\nabla^2 H_{AN} + \frac{1}{2} H H^{AN} \bar \nabla^2 H_{AN} - \frac{1}{8} H^2 \bar \nabla^2 H \nonumber \\
& + \frac{1}{4} H^{MN} H_{MN} \bar \nabla^2 H \label{L0_(3)}
\end{align}
Here, the spacetime indices are raised and lowered by the background metric $\bar G_{MN}$, the covariant derivative $\bar \nabla_M$ is compatible with $\bar G_{MN}$, and $H = \bar G_{MN} H^{MN}$.

Now, we expand the bulk metric perturbation in the number of the boundary data $B_{sab}$
\begin{align}
H_{ab} = \alpha^1 H_{ab}^{(1)} + \alpha^2 H_{ab}^{(2)} + \mathcal{O}(\alpha^3)
\end{align}
where the bookkeeping parameter $\alpha$ is used to count the powers of $B_{\s ab}$. Accordingly, the Dirichlet-type boundary condition \eqref{Hab_bdy_condition} will be implemented as follows:
\begin{align}
& H_{ab}^{(1)}(r\to \infty_\s, \sigma^a) = r^2 \left[ B_{\s ab}(\sigma) + \mathcal{O}(r^{-1}) \right], \nonumber \\
& H_{ab}^{(n\geq2)}(r\to \infty_\s, \sigma^a) = r^2 \left[ 0 + \mathcal{O}(r^{-1}) \right] \label{AdS_bdy_condition_perturb}
\end{align}

Here, we would like to demonstrate that the nonlinear solutions $H_{MN}^{(n\geq2)}$ are unnecessary for the derivation of the EFT action up to cubic order in $B_{\s ab}$. This is mainly due to the implementation of the boundary condition in \eqref{AdS_bdy_condition_perturb}.

Obviously, the background part $S_0^{(0)}$ does not involve any bulk perturbation
\begin{align}
S_0^{(0)} + S_{\rm GH}^{(0)} + S_{\rm ct}^{(0)} = - \int d^4\sigma \, r_h^4\big|_{\infty_2}^{\infty_1}
\end{align}

The linear part $S_0^{(1)}$, combined with $S_{\rm GH}^{(1)}$ and $S_{\rm ct}^{(1)}$, is evaluated as
\begin{align}
S^{(1)} = S_0^{(1)} + S_{\rm GH}^{(1)} + S_{\rm ct}^{(1)} = \int d^4\sigma\, \frac{r_h^4}{2r^2} (3H_{00} + H_{ii})\Big|_{\infty_2}^{\infty_1}
\end{align}
With the boundary condition \eqref{AdS_bdy_condition_perturb}, it is obvious that $H_{MN}^{(n\geq2)}$ does not contribute to the boundary EFT action.

Now, we turn to the quadratic and cubic parts. Practically, the evaluation of them can be simplified by virtue of integration by part
\begin{align}
&S_0^{(2)} + S_0^{(3)} \nonumber \\
= & S_0^{(2)} + S_0^{(3)} - \frac{1}{2} \int d^5\sigma H_{MN} \left[ \frac{\delta S_0^{(2)}}{\delta H_{MN}} +  \frac{\delta S_0^{(3)}}{\delta H_{MN}} \right] \nonumber \\
= & \left[ S_0^{(2)} + S_0^{(3)} - \int d^5\sigma \left( \frac{1}{2} H_{MN} \frac{\delta S_0^{(2)}}{\delta H_{MN}} + \frac{1}{3} H_{MN} \frac{\delta S_0^{(3)}}{\delta H_{MN}} \right) \right] - \frac{1}{6} \int d^5\sigma H_{MN} \frac{\delta S_0^{(3)}}{\delta H_{MN}} \nonumber \\
= & \int d^5\sigma \sqrt{-\bar G} \, \bar \nabla_M t^M - \frac{1}{6} \int d^5\sigma H_{MN} \frac{\delta S_0^{(3)}}{\delta H_{MN}} \label{S0_(2)(3)}
\end{align}
In the second line of \eqref{S0_(2)(3)}, the last term vanishes by the dynamical EOMs \eqref{dynamical_EOMs} and the gauge condition \eqref{gauge_condition_HMN}. The vector $t^M$ is
\begin{align}
t^M =& -\frac{1}{4} H \bar \nabla^{M} H - H^{MN} \bar \nabla^{A} H_{NA} -\frac{1}{2} H_{NA} \bar \nabla^{A} H^{MN} + \frac{3}{4} H_{NA} \bar \nabla^{M} H^{NA} \nonumber \\
& + \frac{1}{4} H \bar \nabla_{N} H^{MN} + \frac{3}{4} H^{MN} \bar \nabla_{N} H + \mathcal{O}(H^3),
\end{align}
which contains both the quadratic and cubic terms in the perturbation $H_{MN}$. Via the Gauss theorem, the total derivative part of \eqref{S0_(2)(3)} is reduced into a surface term
\begin{align}
\int d^5\sigma \sqrt{-\bar G}\, \bar \nabla_M t^M = \int d^4\sigma \sqrt{-\bar \gamma}\, \bar n_M t^M \Big|_{\infty_2}^{\infty_1}
\end{align}
Recall that we only keep the terms up to cubic order in $B_{sab}$. Thus, the contribution from the terms of order $\mathcal{O}(H^3)$ in \eqref{S0_(2)(3)} will be computed by simply substituting $H_{MN}$ by the linearized solution $H_{MN}^{(1)}$. The contribution from the terms of order $\mathcal{O}(H^2)$ in \eqref{S0_(2)(3)} seems to require $H_{MN}^{(2)}$. However, with the boundary condition \eqref{AdS_bdy_condition_perturb}, it is direct to show that such terms actually vanish.

\subsection{Solving linearized dynamical EOMs} \label{linearized_dynamics}

In this section, we study the linearized metric perturbation $H_{MN}^{(1)}$. Admittedly, since the discovery of the AdS/CFT correspondence, the classical dynamics of the linearized gravity in the Schwarzschild-AdS has been extensively studied, for examples see \cite{Policastro:2002se,Policastro:2002tn}. However, as emphasized at several points, in the present work we will take a partially on-shell approach and only solve the dynamical EOMs \eqref{dynamical_EOMs} (or equivalently \eqref{dynamical_EOMs1}).

With the gauge condition \eqref{gauge_condition_HMN}, the only propagating degrees of freedom are in $H_{ab}$. Moreover, the correct set of dynamical EOMs has been identified as \eqref{dynamical_EOMs}. Indeed, for the linearized bulk theory, \eqref{dynamical_EOMs} can be put into a compact form
\begin{align}
E_{ab} \equiv R_{ab}[G] - \frac{1}{2} G_{ab} R[G] - 6 G_{ab} =0 \label{dynamical_EOMs1}
\end{align}
This conclusion simply follows from that the gauge condition \eqref{gauge_fixing_deltaGMN} implies $\delta G^{rM} =0$ but $\delta G^{ab} \neq 0$ for the linearized theory.

We find it more convenient to work with the following bulk variable
\begin{align}
h_{ab}(r,\sigma^a) \equiv r^{-2} H_{ab}^{(1)}(r,\sigma^a),
\end{align}
whose boundary condition inherits from \eqref{AdS_bdy_condition_perturb}
\begin{align}
h_{ab}(r \to \infty_\s, \sigma^a) = B_{\s ab}(\sigma) + \mathcal{O}(r^{-1}). \label{AdS_bdy_condition_hab}
\end{align}
Then, in terms of $h_{ab}$, the dynamical EOMs \eqref{dynamical_EOMs1} are
\begin{align}
0 & = r^5 f(r)\partial_r^2 h_{00} + (5r^4 - 7 r_h^4) \partial_r h_{00} + 2 r^3 \partial_r \partial_0 h_{00} + \frac{8r_h^8}{r^5 f(r)} h_{00} \nonumber \\
& + \frac{3r^2}{f(r)} \left( 1- \frac{3r_h^4}{r^4} \right) \partial_0 h_{00} + \frac{r}{f(r)} \partial_0^2 h_{00} + \frac{2}{3} r \partial_k^2 h_{00} - \frac{4}{3} r \partial_0 \partial_k h_{0k} \nonumber \\
& - \frac{4}{3} r_h^4 f(r) \partial_r h_{kk} - \frac{4r_h^4}{3r^2} \partial_0 h_{kk} + \frac{2}{3} r \partial_0^2 h_{kk} + \frac{1}{3} r f(r) \left( \partial_i^2 h_{kk} -\partial_i \partial_j h_{ij} \right), \nonumber \\
0 & = r^5 \partial_r^2 h_{0i} + 5r^4 \partial_r h_{0i} + \frac{2r^3}{f(r)} \partial_r \partial_0 h_{0i} + \frac{3r^2}{f^2(r)} \left( 1 - \frac{7r_h^4}{3r^4} \right) \partial_0 h_{0i} \nonumber \\
&+ \frac{r}{f^2(r)} \partial_0^2 h_{0i} + \frac{r}{f(r)} \left( \partial_0 \partial_i h_{kk} - \partial_0 \partial_k h_{ki} \right) + \frac{r}{f(r)} \left( {\vec \partial}^{\,2} h_{0i} - \partial_i \partial_k h_{0k} \right), \nonumber \\
0 & = r^5 f(r)\partial_r^2 h_{ij} + (5r^4 - r_h^4) \partial_r h_{ij} + 2r^3 \partial_r \partial_0 h_{ij} + 3r^2 \partial_0 h_{ij} - \frac{2}{3}r_h^4 \delta_{ij} \partial_r h_{kk} \nonumber \\
& - \frac{r}{f(r)} \left( \partial_i \partial_j - \frac{1}{3} \delta_{ij} {\vec\partial}^{\,2} \right) h_{00} + \frac{r}{f(r)} \left( \partial_0 \partial_i h_{0j} + \partial_0 \partial_j h_{0i} - \frac{2}{3} \delta_{ij} \partial_0 \partial_k h_{0k} \right) \nonumber \\
& - \frac{2r_h^4}{3r^2 f(r)} \delta_{ij} \partial_0 h_{kk} + \frac{r}{3f(r)} \delta_{ij} \partial_0^2 h_{kk} + r \left({\vec \partial}^{\,2} h_{ij} - \partial_i \partial_k h_{jk} - \partial_j \partial_k h_{ik} \right. \nonumber \\
& \left. + \frac{1}{3} \delta_{ij} \partial_k \partial_l h_{kl} - \frac{1}{3} \delta_{ij} \vec{\partial}^{\,2}h_{kk} + \partial_i \partial_j h_{kk}  \right), \label{hab_eoms}
\end{align}

Near the AdS boundary $r=\infty_\s$, the linearized metric $h_{ab}$ is expanded as
\begin{align}
h_{ab}(r \to \infty_\s,\sigma^a) = B_{\s ab}(\sigma) + \cdots + \frac{t_{\s ab}(\sigma)}{r^4} + \cdots, \label{hab_bdy_expansion}
\end{align}
where the omitted terms denoted by $\cdots$ are known in terms of the boundary data $B_{\s ab}$. The normalizable mode $t_{\s ab}$ shall be determined by solving \eqref{hab_eoms} over the entire contour of Figure \ref{holographic_SK_contour} and  will be a linear functional of $B_{\s ab}$.

The EOM for $h_{0i}$ looks similar to that of the time-component of a U(1) gauge field in the bulk \cite{Glorioso:2018mmw}. This immediately results in trouble when it comes to imposing two distinct AdS conditions \eqref{AdS_bdy_condition_hab} for $h_{0i}$. This implies that we need an extra condition for $h_{0i}$ at an intermediate location of the contour in Figure \ref{holographic_SK_contour} to ensure that $h_{0i}$ would behave differently at the two AdS boundaries. We follow the treatment of \cite{Glorioso:2018mmw} and impose the following horizon condition
\begin{align}
h_{0i}(r = r_h -\epsilon, \sigma^a) = 0. \label{horizon_condition}
\end{align}

Here, we briefly discuss the physical consequence of the horizon condition \eqref{horizon_condition}. Indeed, it could be viewed as an extra gauge-fixing. With the gauge condition \eqref{gauge_condition_HMN}, we still have residual gauge symmetry \cite{Policastro:2002tn} for the bulk theory, which implies the full diffeomorphism symmetry on the boundary
\begin{align}
\sigma^a \to \sigma^{\prime a} (\sigma). \label{bdy_diffeo}
\end{align}
Thanks to the condition \eqref{horizon_condition}, the full diffeomorphism invariance \eqref{bdy_diffeo} in the fluid spacetime is partially broken to the proposed re-parametrization symmetry \eqref{fluid_reparam}. Interestingly, the recent work \cite{Knysh:2024asf} has explored the relationship between the horizon symmetries and the so-called shift symmetry (or re-parametrization symmetry) for the boundary EFT. A detailed analysis along this line will be presented elsewhere.

 Intriguingly, the time-time component of the bulk metric $h_{00}$ does not satisfy a vanishing horizon condition like \eqref{horizon_condition}, which is essentially linked to the second re-parametrization symmetry in \eqref{fluid_reparam}. Intuitively, this may be understood in the context of linearized hydrodynamics \cite{Kovtun:2012rj}. Take a plane-wave ansatz for the perturbations of energy-momentum densities $\delta T^{0\mu}$. With respect to the spatial wave-vector of the plane-wave, the momentum densities $\delta T^{0i}$ can be decomposed into the transverse component $\delta T^{0\perp}$ and the longitudinal one $\delta T^{0\parallel}$. We know that $\delta T^{0\perp}$ obeys a ``diffusive'' equation, which is the same as a conserved U(1) charge density. At the ideal level, the coupled equations satisfied by $\delta T^{00}$ and $\delta T^{0\parallel}$ could be cast into a ``wave'' equation for $\delta T^{00}$. This observation implies that in the SK-EFT we shall impose the so-called ``chemical shift symmetry'' for the momentum density $\delta T^{0i}$, but we shall not do this for the energy density $\delta T^{00}$. Via the holographic dictionary, $h_{00}$ and $h_{0i}$ correspond to $\delta T^{00}$ and $\delta T^{0i}$, respectively. This analysis immediately motivates the vanishing horizon condition \eqref{horizon_condition} for $h_{0i}$, while no such a condition for $h_{00}$.

With the analogy between the conserved U(1) density and the momentum density $\delta T^{0i}$, the horizon condition \eqref{horizon_condition} may be intuitively interpreted as the presence of a nonzero surface ``momentum'' at the interface $r=r_h - \epsilon$. This is analogous to the situation of an interface in classical electrodynamics. Apparently, this surface momentum arises from the stochastic noise captured by the $a$-variable. Indeed, it is direct to check that the presence of noise renders the radial derivative of $h_{00}$ discontinuous across the surface $r=r_h - \epsilon$.

From the EOMs \eqref{hab_eoms} and the boundary condition \eqref{AdS_bdy_condition_hab}, it is clear that $h_{ab}$ will be a linear functional of the boundary data $B_{\s ab}$. In the hydrodynamic limit, $B_{\s ab}$ is assumed to evolve slowly in the boundary spacetime, which motivates one to solve the EOMs \eqref{hab_eoms} perturbatively. Thus, we expand $h_{ab}$ in terms of the boundary derivative
\begin{align}
h_{ab} = h_{ab}^{[0]} + h_{ab}^{[1]} + \cdots,
\end{align}
where the superscript is used to count the number of the boundary derivatives of $B_{\s ab}$. Accordingly, the AdS boundary condition \eqref{AdS_bdy_condition_hab} will be implemented perturbatively,
\begin{align}
& h_{ab}^{[0]}(r\to \infty_\s, \sigma^a) = B_{\s ab}(\sigma) + \mathcal{O}(r^{-1}), \nonumber \\
& h_{ab}^{[n\geq 1]}(r\to \infty_\s, \sigma^a) = 0 + \mathcal{O}(r^{-1}) \label{AdS_bdy_condition_hab_pert}
\end{align}
Notice that, at the leading order in the boundary derivative expansion, the EOMs for $h_{ab}^{[0]}$ are homogeneous ordinary differential equations, whose generic solutions over the entire contour of Figure \ref{holographic_SK_contour} are easy to work out. Imposing the AdS condition \eqref{AdS_bdy_condition_hab_pert}, we obtain analytical solution for the leading order bulk perturbation $h_{ab}^{[0]}$
\begin{align}
& h_{00}^{[0]}(r) = f(r) B_{\r 00} - \frac{1}{2} \sqrt{f(r)} B_{\a 00} + \frac{1}{6} \left[1-f(r) \right] \sqrt{f(r)} B_{\a ii}, \nonumber \\
& h_{0i}^{[0]}(r) = f(r) B_{\r 0i} - \frac{1}{2} f(r) B_{\a 0i}, \qquad r\in [r_h - \epsilon, \infty_2), \nonumber \\
& h_{0i}^{[0]}(r) = f(r) B_{\r 0i} + \frac{1}{2} f(r) B_{\a 0i}, \qquad r\in [r_h - \epsilon, \infty_1), \nonumber \\
& h_{ij}^{[0]}(r) = B_{\r ij} -  \frac{1}{2} B_{\a ij} - \frac{\rm i}{2\pi} B_{\a ij} \log f(r) + \frac{1}{6} \left[ \frac{\rm i}{\pi} \log f(r) - \sqrt{f(r)} + 1 \right] \delta_{ij} B_{\a kk}. \label{hab_0th}
\end{align}
Here, because of the horizon condition \eqref{horizon_condition}, the solution for $h_{0i}$ becomes piecewise.

We turn to the first order correction in the boundary derivative expansion. The EOMs for $h_{ab}^{[1]}$ are the same as those for $h_{ab}^{[0]}$ except for the source terms, which are built from the leading order solution $h_{ab}^{[0]}$. Here, we report the final solution for $h_{ab}^{[1]}$
\begin{align}
h_{00}^{[1]}(r) = & -2 \zeta(r) \partial_0 h_{00}^{[0]}(r) + \frac{{\rm i} \pi} {4} \left[ f(r) - \sqrt{f(r)} \right] \left( 2 \partial_0 B_{\r 00} + \partial_0 B_{\a 00} \right) \nonumber \\
& + \frac{{\rm i} \pi}{12r_h} \left[1-f(r) \right] \sqrt{f(r)} \left( 2 \partial_0 B_{\r kk} + \partial_0 B_{\a kk} \right), \nonumber \\
h_{0i}^{[1]}(r) = & - 2 \zeta(r) \partial_0 h_{0i}^{[0]}(r), \qquad r\in [r_h - \epsilon, \infty_2), \nonumber \\
h_{0i}^{[1]}(r) = & -2 \zeta(r) \partial_0 h_{0i}^{[0]}(r) + \frac{{\rm i} \pi} {2r_h} f(r) \left(2 \partial_0 B_{\r 0i} + \partial_0 B_{\a 0i} \right), \qquad r\in [r_h - \epsilon, \infty_1), \nonumber \\
h_{ij}^{[1]}(r) = & -2 \zeta(r) \partial_0 h_{ij}^{[0]}(r) + \frac{1}{4r_h} \log f(r)\, (2\partial_0 B_{\r ij} + \partial_0 B_{\a ij}) \nonumber \\
& +\frac{1}{12 r_h} \left[ {\rm i}\pi \left( 1 - \sqrt{f(r)} \right) - \log f(r) \right] \delta_{ij} (2 \partial_0 B_{\r kk} + \partial_0 B_{\a kk} ), \label{hab_1st}
\end{align}
where the function $\zeta(r)$ is
\begin{align}
\zeta(r) &\equiv \int_{\infty_2}^r \frac{dy}{y^2f(y)} = - \frac{1}{4r_h} \left[\pi - 2 \arctan \left(\frac{r}{r_h} \right) - \log(r-r_h) + \log(r+r_h) \right],
\end{align}
with $r$ varying along the entire contour of Figure \ref{holographic_SK_contour}.

We pause to briefly explain how the solutions \eqref{hab_0th} and \eqref{hab_1st} satisfy the relevant UV and horizon boundary conditions. Since $f(r_h)= 0$, the horizon condition \eqref{horizon_condition} is obviously obeyed by $h_{0i}^{[0]}(r)$ and $h_{0i}^{[1]}[r]$ presented in \eqref{hab_0th} and \eqref{hab_1st} once $\epsilon \to 0$ is taken in the end. While $f(r)$ is regular along the entire radial contour, the functions $\sqrt{f(r)}$, $\log f(r)$ and $\zeta(r)$ are multi-valued at the UV boundary
\begin{align}
& \sqrt{f(r)} \xrightarrow{r\to \infty_2} 1- \frac{r_h^4}{2r^4}+ \cdots, \qquad \sqrt{f(r)} \xrightarrow{r\to \infty_1} - \left(1- \frac{r_h^4}{2r^4} \right)+ \cdots, \nonumber \\
& \log f(r) \xrightarrow{r\to \infty_2} 1- \frac{r_h^4}{r^4} + \cdots, \qquad \log f(r) \xrightarrow{r\to \infty_1} 2 {\rm i}\pi + 1- \frac{r_h^4}{r^4} + \cdots, \nonumber \\
& \zeta(r) \xrightarrow{r\to \infty_2} - \frac{1}{r} - \frac{r_h^4}{5r^5} + \cdots, \qquad \zeta(r) \xrightarrow{r\to \infty_1} \frac{{\rm i}\pi}{2r_h}- \frac{1}{r} - \frac{r_h^4}{5r^5} + \cdots,
\end{align}
With this fact in mind, it is straightforward to check that \eqref{hab_0th} and \eqref{hab_1st} satisfy the desired UV boundary conditions \eqref{AdS_bdy_condition_hab_pert}.

We would like to point out that the solutions \eqref{hab_0th} and \eqref{hab_1st} show singular behavior near the event horizon. This reflects the fact that for the bulk metric, both the regular part corresponding to the ingoing mode (dual to the dissipation on the boundary) and the singular part representing the outgoing (Hawking) mode (dual to the fluctuation on the boundary) are consistently kept in a systematic way. This is one of the main advantage of the holographic SK contour prescription of \cite{Glorioso:2018mmw}. Switching off the $a$-type variables, we find that the solutions in \eqref{hab_0th} and \eqref{hab_1st} will reduce to those of \cite{Arnold:2011ja} once the constraint components of the bulk EOMs are imposed.

From the solutions \eqref{hab_0th} and \eqref{hab_1st}, we read off the hydrodynamic expansion of the normalizable modes (cf. \eqref{hab_bdy_expansion})
\begin{align}
t_{1\, 00} = & - \frac{r_h^4}{12} (12 B_{\r 00} + 2 B_{\a kk} + 3 B_{\a 00} )\nonumber \\
& + \frac{{\rm i}\pi}{48} r_h^3 \partial_0 (2 B_{\a kk} - 3 B_{a00} - 4 B_{\r kk} + 6  B_{r00}) + \mathcal{O}(\partial^2), \nonumber \\
t_{2\, 00} = & - \frac{r_h^4}{12}(12  B_{\r 00} - 2 B_{\a kk} - 3 B_{a00} )\nonumber \\
& + \frac{{\rm i} \pi}{48} r_h^3 \partial_0(2 B_{\a kk} - 3 B_{a00} + 4 B_{\r kk} - 6  B_{\r 00}) + \mathcal{O}(\partial^2), \nonumber \\
t_{1\,0i} = & - \frac{r_h^4}{2} ( 2 B_{\r 0i} + B_{\a 0i} ) + \mathcal{O}(\partial^2), \nonumber \\
t_{2\,0i} = & - \frac{r_h^4}{2} ( 2 B_{\r 0i} - B_{\a 0i} ) + \mathcal{O}(\partial^2), \nonumber \\
t_{1\,ij} = & \frac{{\rm i} r_h^4}{2\pi} B_{\a ij} + \frac{r_h^3}{8} \partial_0 (B_{\a ij} - 2 B_{\r ij}) - \delta_{ij} \left\{ \frac{r_h^4}{12\pi} (\pi + 2 {\rm i}) B_{\a kk} \right. \nonumber \\
& \left. - \frac{{\rm i} r_h^3}{48} (\pi + 2{\rm i}) \partial_0 (B_{\a kk} - 2 B_{\r kk}) \right\} + \mathcal{O}(\partial^2), \nonumber \\
t_{2\,ij} = & \frac{{\rm i} r_h^4}{2\pi} B_{\a ij} - \frac{r_h^3}{8} \partial_0 (B_{\a ij} + 2 B_{\r ij}) + \delta_{ij} \left\{ \frac{r_h^4}{12\pi} (\pi - 2 {\rm i}) B_{\a kk} \right. \nonumber \\
& \left. + \frac{{\rm i} r_h^3}{48} (\pi - 2{\rm i}) \partial_0 (B_{\a kk} + 2 B_{\r kk}) \right\} + \mathcal{O}(\partial^2).
\end{align}

\subsection{Holographic effective action} \label{holo_action}

As shown in section \ref{holo_RG}, in the saddle point approximation, the boundary effective action is simply given by the partially on-shell bulk action
\begin{align}
S_{eff} = S_0|_{\rm p.o.s.} + S_{\rm GH} + S_{\rm ct} \label{Seff_S}
\end{align}
With the linearized solutions presented in section \ref{linearized_dynamics}, we are ready to compute \eqref{Seff_S}. In accord with \eqref{S0_expansion}, we expand $S_{eff}$ similarly
\begin{align}
S_{eff} = S_{eff}^{(0)} + S_{eff}^{(1)} + S_{eff}^{(2)} + S_{eff}^{(3)} + \cdots.
\end{align}
With the bulk Lagrangian \eqref{L0_(0)}-\eqref{L0_(2)}, we advance by simplifying the bulk action a bit using the bulk EOMs. Up to quadratic order, \eqref{Seff_S} can be reduced to a surface term\footnote{Here, we have adopted the treatment of the second line of \eqref{S0_(2)(3)} for $S_0^{(2)}$. Moreover, divergences beyond quadratic order in boundary spacetime derivative are omitted near the AdS boundary.}
\begin{align}
& S_{eff}^{(0)} + S_{eff}^{(1)} + S_{eff}^{(2)}= \int d^4\sigma \left\{ - r_h^4 + \frac{r_h^4}{2} \left( 3 h_{00} + h_{ii} \right) \right. \nonumber \\
& \left. + \frac{r_h^4}{8} \left(3 h_{00} h_{00} + 10 h_{00} h_{ii} - 12 h_{0i} h_{0i} + h_{ii} h_{jj}  - 2 h_{ij} h_{ij} \right) \right. \nonumber \\
& \left.  + \frac{r^5}{8} \left[ 2 \left( h_{0i} h_{0i} \right)' -2 \left(h_{00} h_{ii} \right)' + f(r) \left( h_{ii} h_{jj} \right)' - f(r) \left( h_{ij} h_{ij} \right)' \right] \right\} \bigg|_{\infty_2}^{\infty_1} \label{Seff_012_bulk}
\end{align}
where a prime denotes the radial derivative.

Now, we present our results for the boundary effective action. From \eqref{Seff_012_bulk}, the background part $S_{eff}^{(0)}$ is
\begin{align}
S_{eff}^{(0)} = \int d^4\sigma (-r_h^4)\big|_{\infty_2}^{\infty_1} = 0 \label{Seff_0}
\end{align}

From \eqref{Seff_012_bulk}, the linear part $S_{eff}^{(1)}$ is computed as
\begin{align}
S_{eff}^{(1)} & = \int d^4\sigma\, \frac{r_h^4}{2} \left( 3 h_{00} + h_{ii} \right) \bigg|_{r=\infty_2}^{r=\infty_1} \nonumber \\
& = \int d^4\sigma \, \frac{r_h^4}{2} \left[ 3 B_{1\,00} + B_{1\,ii} \right] - \int d^4\sigma \, \frac{r_h^4}{2} \left[ 3 B_{2\,00} + B_{2\,ii} \right] \nonumber \\
& = \int d^4\sigma \, \frac{r_h^4}{2} \left[ 3 B_{\a 00} + B_{\a ii} \right]. \label{Seff_1}
\end{align}

From \eqref{Seff_012_bulk}, the quadratic part $S_{eff}^{(2)}$ is
\begin{align}
S_{eff}^{(2)} = & \int d^4\sigma \left(-2 B_{1\,0i} t_{1\,0i} + B_{1\,00} t_{1\,ii} + B_{1\,ii} t_{1\,00} - B_{1\,ij} t_{1\,ij} + B_{1\,ii} t_{1\,jj} \right) \nonumber \\
- & \int d^4\sigma \left(-2 B_{2\,0i} t_{2\,0i} + B_{2\,00} t_{2\,ii} + B_{2\,ii} t_{2\,00} - B_{2\,ij} t_{2\,ij} + B_{2\,ii} t_{2\,jj} \right) \nonumber \\
=& \int d^4\sigma\, \left\{ \frac{3r_h^4}{4} B_{\a 00} B_{\r 00} + r_h^4 B_{\a 0i} B_{\r 0i} - \frac{r_h^4}{4} B_{\a ii} B_{\r 00}  + \frac{3r_h^4}{4} B_{\a 00} B_{\r ii} \right. \nonumber \\
& \qquad \qquad 
+ \left. \frac{r_h^4}{4} B_{\a ii} B_{\r jj} 
- \frac{r_h^4}{2} B_{\a ij} B_{\r ij} + \frac{{\rm i} r_h^4}{6 \pi} \left( 3 B_{\a ij} B_{\a ij} - B_{\a ii} B_{\a jj} \right) \right. \nonumber \\
& \qquad \qquad \left. + \frac{r_h^3}{6} \left( B_{\a ii} \partial_0 B_{\r jj} - 3 B_{\a ij} \partial_0 B_{\r ij} \right) - \frac{{\rm i}\pi r_h^3}{8} B_{\a 00}\partial_0 B_{\a ii} \right\}.  \label{Seff_2}
\end{align}

Finally, we turn to the cubic part $S_{eff}^{(3)}$. From \eqref{S0_(2)(3)}, we consider
\begin{align}
\hat S_3^{(0)} = S_{0}^{(3)} - \frac{1}{2} H_{MN} \frac{\delta S_0^{(3)}}{\delta H_{MN}}
\end{align}
Then, we have
\begin{align}
S_{eff}^{(3)} = \hat S_3^{(0)} + S_{\rm GH}^{(3)} + S_{\rm ct}^{(3)}
\end{align}
which is computed as
\begin{align}
S_{eff}^{(3)} = & \int d^4 \sigma\, r_h^4 \left\{ \frac{9}{16} B_{\a 00} B_{r00}^2 - \frac{1}{16} B_{\a ii} B_{\r 00}^2 + \frac{3}{8} B_{\a 00} B_{\r 00} B_{\r ii} - \frac{3}{4} B_{\a 00} B_{\r 0i}^2 \right. \nonumber \\
& \left. + \frac{5}{2} B_{\a 0i} B_{\r 0i} B_{\r 00} - \frac{7}{24} B_{\a ii} B_{\r jj} B_{\r 00} + \frac{3}{16} B_{\a 00} B_{\r ii}^2 + \frac{9}{4} B_{\a ii} B_{\r 0k}^2 \right. \nonumber \\
& \left. + \frac{5}{2} B_{\a 0i} B_{\r 0i} B_{\r jj} - \frac{1}{4} B_{\a ij} B_{\r ij} B_{\r 00} - \frac{3}{8} B_{\a 00} B_{\r ij} B_{\r ij} - \frac{5}{2} B_{\a ij} B_{\r 0i} B_{\r 0j} \right. \nonumber \\
& \left. - 5 B_{\a 0i} B_{\r 0j} B_{\r ij} - \frac{13}{48} B_{\a ii} B_{\r jj}^2  + \frac{3}{8} B_{\a ii} B_{\r kl} B_{\r kl} + \frac{1}{4} B_{\a ij} B_{\r ij} B_{\r kk} \right. \nonumber \\
& \left. - \frac{1}{2} B_{\a ik} B_{\r jk} B_{\r ij} + \frac{{\rm i}}{\pi} \left[ \frac{1}{3} B_{\a ii}^2 B_{\r 00} + \frac{1}{4} B_{\a 00} B_{\a ii} B_{\r jj} - \frac{4}{3} B_{\a ii} B_{\a 0j} B_{\r 0j} \right. \right. \nonumber \\
& \left. \left.  - B_{\a ij} B_{\a ij} B_{\r 00} - \frac{3}{4} B_{\a 00} B_{\a ij} B_{\r ij} + 4 B_{\a ij} B_{\a 0i} B_{\r 0j} - \frac{31}{36} B_{\a ii} B_{\a jj} B_{\r kk} \right. \right. \nonumber \\
& \left. \left.  + B_{\a ij} B_{\a ij} B_{\r kk} + \frac{35}{12} B_{\a ii} B_{\a jk} B_{\r jk} - 4 B_{\a ik} B_{\a jk} B_{\r ij} \right] + \frac{3}{64} B_{\a 00}^3 \right. \nonumber \\
& \left.  + \frac{3}{64} B_{\a 00}^2 B_{\a ii} - \frac{3}{16} B_{\a 00} B_{\a 0i}^2 + \frac{8-7 \pi^2}{96 \pi^2} B_{\a 00} B_{\a ii} B_{\a jj} + \frac{19}{48} B_{\a 0i}^2 B_{ajj} \right. \nonumber \\
& \left. - \frac{4 + 3 \pi^2}{32 \pi^2}  B_{\a 00} B_{\a ij} B_{\a ij} - \frac{1}{8} B_{\a 0i} B_{\a 0j} B_{\a ij} + \frac{24 - 91 \pi^2}{1728 \pi^2} B_{\a ii}^3 \right. \nonumber \\
& \left. + \frac{9 \pi^2 - 4}{96 \pi^2} B_{\a ii} B_{\a jk} B_{\a jk} - \frac{1}{24} B_{\a ij} B_{\a ik} B_{\a jk} \right\}.  \label{Seff_3}
\end{align}

The equations \eqref{Seff_0}, \eqref{Seff_1}, \eqref{Seff_2} and \eqref{Seff_3} stand for one of our main results. For simplicity, we have suppressed the coordinate $\sigma^a$ in all $B$'s and $t$'s above. We have checked that the holographic action \eqref{Seff_1}, \eqref{Seff_2} and \eqref{Seff_3} satisfy all the symmetry requirements listed in \eqref{Z2_reflection}-\eqref{dynamical_KMS}. Due to the linearization approximation \eqref{X_g_linearization} and \eqref{B12_expand}, we find that the fluid re-parametrization symmetry \eqref{fluid_reparam} is valid up to certain nonlinear terms.

However, the results \eqref{Seff_1}, \eqref{Seff_2} and \eqref{Seff_3} look a bit obscure. This has to do with the following two facts. First, the dependence of $B_{\s ab}(\sigma)$ on the external source and the dynamical field is implicit. Second, we work in the fluid spacetime spanned by $\sigma^a$. To circumvent this shortcoming, we will reformulate the results \eqref{Seff_1}, \eqref{Seff_2} and \eqref{Seff_3} in the physical spacetime by exploiting the linearization expansion in \eqref{X_g_linearization} and \eqref{B12_expand}. Accordingly, the results \eqref{Seff_1}, \eqref{Seff_2} and \eqref{Seff_3} will be re-organized by the powers of the external source $A_{\s\mu\nu}(\sigma)$ and the dynamical field $\pi_{\s\mu}(\sigma)$. Here, owing to the truncation made in \eqref{B12_expand}, it is then valid to track the terms up to quadratic order only for the holographic action.

We know that in the fluid spacetime, the dynamical variables are $X_1^\mu(\sigma^a)$ and $X_2^\mu(\sigma^a)$. Equivalently, in the $\r\a$-basis we have
\begin{align}
X_\r^\mu(\sigma^a)\equiv \frac{1}{2}\left[ X_1^\mu(\sigma^a) + X_2^\mu(\sigma^a) \right], \qquad X_\a^\mu(\sigma^a)\equiv X_1^\mu(\sigma^a) - X_2^\mu(\sigma^a),
\end{align}
which, in the linearization approximation \eqref{X_g_linearization}, reads
\begin{align}
X_\r^\mu(\sigma^a) = \delta_a^\mu \sigma^a + \pi_\r^\mu(\sigma), \qquad X_\a^\mu(\sigma^a) = \pi_\a^\mu(\sigma). \label{Xr_Xa_linear}
\end{align}
However, when it comes to studying physical observables such as the correlation functions, it is more convenient to rewrite the action integral in the physical spacetime in which the external source $g_{\s\mu\nu}(x)$ is defined. Following \cite{Crossley:2015evo}, the physical spacetime is spanned by the coordinate $x^\mu$ defined as
\begin{align}
x^\mu \equiv X_\r^\mu(\sigma^a)
\end{align}
which would be viewed as a field in the fluid spacetime, i.e., $x^\mu = x^\mu(\sigma^a)$. Inverting this relation, we have
\begin{align}
\sigma^a = \sigma^a(x^\mu)
\end{align}
Meanwhile, we can also view $X_\a^\mu$ as a field in $x^\mu$-coordinate, i.e., $X_\a^\mu = X_\a^\mu(x^\mu)$. Therefore, in the physical spacetime, the dynamical variables are $\sigma^a(x^\mu)$ and $X_\a^\mu(x^\mu)$.

In the linearization approximation \eqref{X_g_linearization}, the first relation in \eqref{Xr_Xa_linear} can be inverted perturbatively
\begin{align}
\delta_a^\mu \sigma^a =  x^\mu - \pi_\r^\mu(x) + \pi_\r^\nu(x) \partial_\nu \pi_\r^\mu(x) + \cdots
\end{align}
Thus, in the physical spacetime we have the dynamical variables $\pi_\r^\mu(x)$ and $\pi_\a^\mu(x)$. The differential volume element changes as
\begin{align}
d^4\sigma =\det\left[ \frac{\partial \sigma^a}{\partial x^\mu} \right] d^4x = d^4x\, \left[1 - \partial_\mu \pi_\r^\mu(x) + \mathcal{O}(\pi^2) \right] \label{measure_Jacobian}
\end{align}
where the terms beyond quadratic order in $\pi$ are not needed for capturing quadratic order terms in the EFT action. In addition, we need
\begin{align}
&A_{\a ab}(\sigma) = \delta_a^\mu \delta_b^\nu \left[ A_{\a \mu\nu}(x) - \pi_\r^\alpha(x) \partial_\alpha A_{\a \mu\nu}(x) \right] + \cdots , \nonumber \\
&\partial_a \pi_{\r,\a}^\alpha(\sigma) = \delta_a^\mu \partial_\mu \pi_{\r,\a}^\alpha(x) \label{A_pdpi_sigma_x}
\end{align}

With these issues clarified, we are ready to rewrite the results \eqref{Seff_1}, \eqref{Seff_2} and \eqref{Seff_3} in the physical spacetime. Accordingly, we re-organize the EFT action by the powers of the external sources $A_{\r\mu\nu}(x)$, $A_{\a\mu\nu}(x)$ and the dynamical fields $\pi_\r^\mu(x)$, $\pi_\a^\mu(x)$. Schematically, we have
\begin{align}
S_{eff} = S_s + S_{sd} + S_d =\int d^4x \left[ \mathcal L_s + \mathcal L_{sd} + \mathcal L_d \right] \label{separation1}
\end{align}
where $\mathcal L_s$ stands for the source part containing the external source only, $\mathcal L_{sd}$ denotes the crossing terms having both the external source and the dynamical field, and $\mathcal L_d$ represents the terms with the dynamical field only. Below we present the results.

The source part $\mathcal L_s$ is
\begin{align}
\mathcal L_s = & \frac{3}{2}r_h^4 A_{\a 00}(x) + \frac{1}{2}r_h^4 A_{\a ii}(x) + \frac{3r_h^4}{4} A_{\a 00} A_{\r 00} + r_h^4 A_{\a 0i} A_{\r 0i} - \frac{r_h^4}{4} A_{\a ii} A_{\r 00} \nonumber \\
& + \frac{3r_h^4}{4} A_{\a 00} A_{\r ii} + \frac{r_h^4}{4} A_{\a ii} A_{\r jj} - \frac{r_h^4}{2} A_{\a ij} A_{\r ij} + \frac{{\rm i} r_h^4}{6 \pi} \left( 3 A_{\a ij} A_{\a ij} - A_{\a ii} A_{\a jj} \right) \nonumber \\
& + \frac{r_h^3}{6} \left( A_{\a ii} \partial_0 A_{\r jj} - 3 A_{\a ij} \partial_0 A_{\r ij} \right) - \frac{{\rm i}\pi r_h^3}{8} A_{\a 00}\partial_0 A_{\a ii} \label{Ls}
\end{align}
The crossing part $\mathcal L_{sd}$ is
\begin{align}
\mathcal L_{sd} = & - \frac{3}{2}r_h^4 A_{\a 00} \partial_\mu \pi_\r^\mu - \frac{1}{2}r_h^4 A_{\a ii} \partial_\mu \pi_\r^\mu - 4 r_h^4 A_{\r 0i} \partial^0 \pi_\a^i  - 4 r_h^4 A_{\a 0i} \partial^0 \pi_\r^i\nonumber \\
& + \frac{1}{3} r_h^3 \partial_{0} A_{\r ii} \partial_j \pi_\a^j - \frac{1}{3} r_h^3 \partial_{0} A_{\a ii} \partial_j \pi_\r^j  - r_h^3 \partial_{0} A_{\r ij} \partial^i \pi_\a^j + r_h^3 \partial_{0} A_{\a ij} \partial^i \pi_\r^j \nonumber \\
& + \frac{\rm i}{\pi} r_h^4 A_{\a ij} \partial^i \pi_\a^j - \frac{2 \rm i}{3 \pi} r_h^4  A_{\a ii} \partial_j \pi_\a^j - \frac{{\rm i} \pi r_h^3}{4} \partial_{0} A_{\a 00} \partial_i \pi_\a^i - \frac{{\rm i} \pi r_h^3}{4} \partial_{0} A_{\a ii} \partial_0 \pi_\a^0  \label{Lsd}
\end{align}
The dynamical part $\mathcal L_d$ is
\begin{align}
\mathcal L_d = & 3 r_h^4 \partial_0 \pi_{\r 0} \partial^0 \pi_\a^0 - 4 r_h^4 \partial_0 \pi_{\r i} \partial^0 \pi_\a^i - r_h^4 \partial_0 \pi_\r^0 \partial_i \pi_\a^i - r_h^4 \partial_0 \pi_\a^0 \partial_i \pi_\r^i - r_h^4 \partial_i \pi_\r^i \partial_j \pi_\a^j   \nonumber \\
& - \frac{r_h^3}{3} \partial_0 \partial_i \pi_\r^i \partial_j \pi_\a^j - r_h^3 \partial_0 \partial_i \pi_{\r j} \partial^i \pi_\a^j + \frac{{\rm i} r_h^4}{3 \pi} \partial_i \pi_\a^i \partial_j \pi_\a^j + \frac{{\rm i} r_h^4}{\pi} \partial_i \pi_{\a j} \partial^i \pi_\a^j \nonumber \\
& + \frac{{\rm i} r_h^3}{2} \partial_0 \partial_0 \pi_\a^0 \partial_i \pi_\a^i  \label{Ld}
\end{align}
Here, we have suppressed $x^\mu$ in all $A$'s and $\pi$'s of \eqref{Ls}-\eqref{Ld}.  To be consistent with the truncations made in \eqref{Bra_expand}, \eqref{measure_Jacobian} and \eqref{A_pdpi_sigma_x}, we have tracked the terms up to quadratic order only in \eqref{Ls}-\eqref{Ld}. However, our holographic results \eqref{Seff_1}, \eqref{Seff_2} and \eqref{Seff_3} are sufficient in capturing cubic terms omitted in \eqref{Ls}-\eqref{Ld}, which would require to add more nonlinear terms in the expansion for $B_{\s ab}$ in \eqref{B12_expand} and is obviously tedious. This study will be left as a future project.

\subsection{Hydrodynamic modes and correlation functions from EFT} \label{modes_correlation_function}

In this section, we use the holographic results derived in section \ref{holo_action} to study the hydrodynamic modes and correlation functions of the boundary stress tensor. The basic goal is to provide further support on the correctness of our results. This will produce the relevant results obtained previously, e.g., \cite{Policastro:2002tn,Kovtun:2005ev,Arnold:2011ja}.

\subsubsection{Hydrodynamic modes}

First, we consider the hydrodynamic modes predicted by the holographic EFT of section \ref{holo_action}. To this end, we consider the dynamical EOM for $\pi_{\r\mu}(x)$, which is obtained from the variation of $S_{eff}$ with respect to the auxiliary variable $\pi_{\a\mu}(x)$
\begin{align}
\frac{\delta S_{eff}}{\delta \pi_{\a\mu}(x)} = 0 \label{eom_pir}
\end{align}
To get the dispersion relations for the dynamical modes, it is sufficient to switch off the external metrics $A_{\a\mu\nu}(x)$, $A_{\r\mu\nu}(x)$ and the auxiliary variable $\pi_{\a\mu}(x)$ in the dynamical EOM \eqref{eom_pir}, i.e., we just need to focus on the dynamical part $\mathcal L_d$ of \eqref{Ld}. Then, in the Fourier space, \eqref{eom_pir} leads to the dispersion equations
\begin{align}
&{\rm shear ~~ channel}:~~~ \tilde \omega + \frac{1}{2} {\rm i} \tilde k^2 + \cdots = 0, \nonumber \\
&{\rm sound ~~ channel}:~~~ \tilde \omega^2 - \frac{1}{3} \tilde k^2 + {\rm i} \frac{2}{3} \tilde \omega \tilde k^2 + \cdots = 0,
\end{align}
which give correct shear mode and sound mode \cite{Policastro:2002tn,Kovtun:2005ev}
\begin{align}
&{\rm shear ~~ mode}:~~~ \tilde \omega = - \frac{1}{2} {\rm i} \tilde k^2 + \cdots, \nonumber \\
&{\rm sound ~~ mode}:~~~ \tilde \omega = \pm  \frac{1}{\sqrt{3}}\tilde k - \frac{2}{3} {\rm i} \tilde k^2 + \cdots. \label{shear_sound_dispersion}
\end{align}
Here, we used the dimensionless four-momentum $\tomega \equiv \omega / (2 r_h) ,\tk \equiv k / (2 r_h) $.

\subsubsection{Stochastic hydrodynamics recovered}

Within the hydrodynamic EFT, the conserved stress tensor is defined as \cite{Crossley:2015evo}
\begin{align}
T_1^{\mu\nu}(x) \equiv \frac{2}{\sqrt{-g_1(x)}}\frac{\delta S_{eff}}{\delta g_{1\mu\nu}(x)}, \qquad \qquad T_2^{\mu\nu}(x) \equiv - \frac{2}{\sqrt{-g_2(x)}} \frac{\delta S_{eff}}{\delta g_{2\mu\nu}(x)}
\end{align}
which are actually the off-shell energy-momentum tensor associated with the upper and lower branches of the SK closed time path. The physical part of the energy-momentum tensor is
\begin{align}
T_\r^{\mu\nu}(x) &= \frac{1}{2} \left[ T_1^{\mu\nu}(x) + T_2^{\mu\nu}(x) \right] \nonumber \\
& = \frac{2}{\sqrt{-g_\r}} \frac{\delta S_{eff}}{\delta g_{\a \mu\nu}(x)} \nonumber \\
& = [2 - \eta^{\mu\nu} A_{\r \mu\nu}(x)] \frac{\delta S_{eff}}{\delta A_{\a \mu\nu}(x)}
\end{align}
where in the second equality we have ignored the terms beyond the linear order in $\a$-variables and in the last equality we have utilized the linearization approximation \eqref{X_g_linearization}. Then, the physical stress tensor can be further split into the hydrodynamic and stochastic parts \cite{Crossley:2015evo}
\begin{align}
T_\r^{\mu\nu} = T_{\rm hydro}^{\mu\nu} + T_{\rm stoc}^{\mu\nu}
\end{align}
Here, $T_{\rm hydro}^{\mu\nu}$ does not contain any $\a$-variables and thus is the hydrodynamic stress tensor; and $T_{\rm stoc}^{\mu\nu}$ contains one $\a$-variable and represents the stochastic force. Apparently, $T_{\rm hydro}^{\mu\nu}$ will be a functional of the dynamical variable $\pi_\r^\mu(x)$ and the external source $A_{\r \mu\nu}$
\begin{align}
T_{\rm hydro}^{00} &= 3r_h^4 - 3r_h^4 \partial_\mu \pi_\r^\mu + 3r_h^4 A_{\r 00}, \nonumber \\
T_{\rm hydro}^{0i} &= T_{\rm hydro}^{i0} = r_h^4 A_{\r 0i} - 4r_h^4\partial^0 \pi_\r^i, \nonumber \\
T_{\rm hydro}^{ij} &= r_h^4 \delta_{ij} - r_h^4 A_{\r ij} - r_h^4 \delta_{ij} \partial_\mu \pi_\r^\mu + \frac{1}{3}r_h^3 \delta_{ij} \partial_0 A_{\r kk} - r_h^3 \partial_0 A_{\r ij} \nonumber \\
& - r_h^3 \partial_0 \left(\partial^i \pi_\r^j + \partial^j \pi_\r^i \right) + \frac{2}{3} r_h^3 \delta_{ij} \partial_0 \partial_k \pi_\r^k \label{T_hydro}
\end{align}
Meanwhile, the stochastic part reads
\begin{align}
T_{\rm stoc}^{00} &= \frac{{\rm i} \pi r_h^3}{4} \partial_0 \partial_i \pi_\a^i - \frac{{\rm i} \pi r_h^3}{8} \partial_0 A_{\a ii}, \nonumber \\
T_{\rm stoc}^{0i} &= T_{\rm stoc}^{i0}= 0, \nonumber \\
T_{\rm stoc}^{ij} &= \frac{{\rm i} }{2\pi}r_h^4 \left(\partial^i \pi_\a^j + \partial^j \pi_\a^i \right) - \frac{2\rm i}{3\pi}r_h^4 \delta_{ij} \partial_k \pi_\a^k + \frac{{\rm i}\pi r_h^3}{4} \delta_{ij} \partial_0^2 \pi_\a^0 \nonumber \\
& + \frac{{\rm i} r_h^4}{\pi} A_{\a ij} - \frac{{\rm i} r_h^4}{3\pi} \delta_{ij} A_{\a kk} + \frac{{\rm i} \pi r_h^3}{8} \delta_{ij} \partial_0 A_{\a 00}
\end{align}

Indeed, the EFT Lagrangian could be schematically written as \cite{Crossley:2015evo}
\begin{align}
\mathcal L = \pi_{\a \nu} \partial_\mu T_{\rm hydro}^{\mu\nu} + \frac{\rm i}{2} \pi_{\a \mu} \mathcal M^{\mu\nu} \pi_{\a \nu} + \mathcal L_s \label{L_schematic}
\end{align}
where
\begin{align}
&\mathcal M^{00}=0, \qquad \mathcal M^{0i}= \frac{1}{2} r_h^3 \partial_0^2 \partial_i, \qquad  \mathcal M^{i0}= - \frac{1}{2} r_h^3 \partial_0^2 \partial_i, \nonumber \\
&\mathcal M^{ij} = - \frac{2r_h^4}{3\pi} \partial_i \partial_j - \frac{2r_h^4}{\pi} \delta_{ij}{\vec\partial}^{\,2}
\end{align}
The dynamical EOM \eqref{eom_pir} gives the conservation law
\begin{align}
\partial_\mu T_\r^{\mu\nu} =0 \Longrightarrow \partial_\mu T_{\rm hydro}^{\mu\nu} = \zeta^\nu
\end{align}
where $\zeta^\nu$ is the stochastic noise, whose distribution is governed by those terms quadratic in $\pi_\a^\mu$ in the Lagrangian. Then, we have the thermal noise $\zeta^\mu$ obeying the Gaussian distribution \cite{Crossley:2015evo}
\begin{align}
\langle \zeta^\mu(x)\rangle =0, \qquad \qquad \langle \zeta^\mu(x) \zeta^\nu(x') \rangle = \mathcal M^{\mu\nu} \delta^{(4)}(x-x') \label{noise}
\end{align}
To first order in the derivative expansion, we can ignore $\mathcal M^{0i}$- and $\mathcal M^{i0}$-terms in the Lagrangian \eqref{L_schematic}. Accordingly, \eqref{noise} reduces to that of the stochastic model for the first order relativistic hydrodynamics \cite{Kovtun:2012rj}. However, for generic case, the EFT prediction \eqref{noise} will naturally go beyond the stochastic model treatment.

In order to cast $T_{\rm hydro}^{\mu\nu}$ of \eqref{T_hydro} into standard hydrodynamic form \eqref{hydro_const_relation}, we need to replace $\pi_\r^\mu$ by the fluid velocity fields \cite{Crossley:2015evo}
\begin{align}
u_\s^\mu \equiv \frac{1}{b_\s} \frac{\partial X_\s^\mu}{\partial \sigma^0}, \qquad {\rm with}~~ b_\s \equiv \sqrt{- g_{\s \mu\nu} (X_\s^\mu) \frac{\partial X_\s^\mu}{\partial \sigma^0} \frac{\partial X_\s^\nu}{\partial \sigma^0}}
\end{align}
The physical fluid velocity is defined as
\begin{align}
u^\mu(x) \equiv \frac{1}{2}(u_1^\mu + u_2^\mu)
\end{align}
which, in the linearization approximation \eqref{X_g_linearization}, is linearized as
\begin{align}
u^0(x) = 1 + \frac{1}{2} A_{\r 00}(x) + \cdots, \qquad \qquad u^i(x) = \partial_0 \pi_\r^i(x) + \cdots
\end{align}

Eventually, we successfully rewrite the result \eqref{T_hydro} into the following standard hydrodynamic form\footnote{According to the linearization approximation \eqref{X_g_linearization}, the inverse metric $g_\s^{\mu\nu}$ is linearized as $g_\s^{\mu\nu} = \eta^{\mu\nu} + A_\s^{\mu\nu} = \eta^{\mu\nu} - \eta^{\mu\alpha} \eta^{\nu \beta} A_{\s\alpha\beta} + \cdots$.} (cf. the equation \eqref{hydro_const_relation})
\begin{align}
T_{\rm hydro}^{\mu\nu} = \epsilon u^\mu u^\nu + P (g_\r^{\mu\nu} + u^\mu u^\nu) - \eta_0 \sigma^{\mu\nu}
\end{align}
where $\sigma^{\mu\nu}$ is the shear tensor
\begin{align}
\sigma^{\mu\nu} \equiv \nabla^\mu u^\nu + \nabla^\nu u^\mu - \frac{2}{3}(u^\mu u^\nu + g_\r^{\mu\nu}) \nabla_\alpha u^\alpha
\end{align}
The holographic value for the shear viscosity is $\eta_0 = r_h^3 = (\pi T)^3$. Interestingly, the fluid's energy density $\epsilon$ and pressure $p$ can be split as
\begin{align}
\epsilon = \epsilon_0 + \delta \epsilon, \qquad P = P_0 + \delta P
\end{align}
where $\epsilon_0 = 3 P_0 = 3r_h^4 = 3 (\pi T)^4$ are the equilibrium counterparts while $\delta \epsilon$ and $\delta P$ denote the non-equilibrium corrections
\begin{align}
\delta \epsilon = - 3 \partial_\mu \pi_\r^\mu, \qquad \qquad \delta P = - \partial_\mu \pi_\r^\mu
\end{align}

\subsubsection{Two-point correlators}

We turn to computing the stress tensor's two-point correlators based on the holographic effective action. With \eqref{separation1}, the generating functional \eqref{Z_EFT} is
\begin{align}
Z[A_{\r\mu\nu}, A_{\a\mu\nu}] = e^{{\rm i} S_s} \int [D \pi_{\r\mu}] [D \pi_{\a\mu}] e^{{\rm i} S_{sd} + {\rm i} S_d} \label{Z_Ss_Ssd_Sd}
\end{align}
which yields the full set of two-point correlators
\begin{align}
G_{\r\a}^{\mu\nu|\rho\sigma}(K) &= \frac{2\,\delta^2 W}{\delta A_{\a\mu\nu}(-K) \delta A_{\r\rho\sigma}(K)}, \qquad
G_{\a\r}^{\mu\nu|\rho\sigma}(K) = \frac{2\,\delta^2 W}{\delta A_{\r\mu\nu}(-K) \delta A_{\a\rho\sigma}(K)}, \nonumber \\
G_{\r\r}^{\mu\nu|\rho\sigma}(K) &= \frac{2\,\delta^2 W}{{\rm i}\, \delta A_{\a\mu\nu}(-K) \delta A_{\a\rho\sigma}(K)}, \label{G_RAS}
\end{align}
where $W \equiv -{\rm i} \log Z$ is the generating functional for the connected correlators.

However, the separation made in \eqref{separation1} has a drawback: the source part $S_s$ is not invariant under the diffeomorphism transformation of the external metric
\begin{align}
\delta A_{\s\mu\nu} = \partial_\mu \xi_{\s\nu} + \partial_\nu \xi_{\s\mu} + \xi_\s^\lambda \partial_\lambda A_{\s\mu\nu} + A_{\s \lambda \mu} \partial_{\nu} \xi^{\lambda}_{\s} + A_{\s \lambda \nu} \partial_{\mu} \xi^{\lambda}_{\s} + \dots, \label{diffeo_source}
\end{align}
where $\xi_{\s\mu}$ is an arbitrary infinitesimal field generating the diffeomorphism transformation \eqref{diffeo_source}. To overcome this shortcoming, instead of \eqref{separation1}, we are motivated to separate the effective action as follows \cite{Crossley:2015evo}
\begin{align}
S_{eff} = S_{\rm inv} + \tilde S_{eff},  \label{separation2}
\end{align}
where $S_{\rm inv}$ is invariant under \eqref{diffeo_source}. Apparently, $S_{\rm inv}$ shall contain the external source only. In addition, we find it more convenient to work with the Fourier modes defined by
\begin{align}
A_{\s\mu\nu}(x) = \int \frac{d^4K}{(2\pi)^4} A_{\s\mu\nu}(K)e^{{\rm i} K_\mu x^\mu}, \qquad \pi_{\s\mu}(x) = \int \frac{d^4K}{(2\pi)^4} \pi_{\s\mu}(K)e^{{\rm i} K_\mu x^\mu},
\end{align}
where, without loss of generality, we choose $K_\mu = (-\omega, k, 0, 0)$. In the Fourier space, the invariant part $S_{\rm inv}$ and the remaining part $\tilde S_{eff}$ are collected in the appendix \ref{cal_detail}, see \eqref{S_inv} and \eqref{tilde_Seff} for details.

Based on the separation \eqref{separation2}, the partition function \eqref{Z_EFT} is computed as
\begin{align}
Z[A_{\r\mu\nu}, A_{\a\mu\nu}] = e^{{\rm i} S_{\rm inv}[A_{\mu\nu}] + {\rm i} \tilde W[A_{\mu\nu}]} = e^{{\rm i} S_{\rm inv}[A_{\mu\nu}]} \int [D \pi_{\r\mu}] [D \pi_{\a\mu}] e^{{\rm i} \tilde S_{eff}} \label{Z_Sinv_tilde_S}
\end{align}
so that the generating functional for the connected correlators is
\begin{align}
W \equiv -{\rm i} \log Z = S_{\rm inv} + \tilde W \label{W_Sinv_Wt}
\end{align}
Performing the path integral over the dynamical field in \eqref{Z_Sinv_tilde_S}, we obtain
\begin{align}
\tilde W = \tilde W_{\rm tensor} + \tilde W_{\rm vector} + \tilde W_{\rm scalar} \label{tilde_W}
\end{align}
where various parts can be found in \eqref{W_tensor}-\eqref{W_scalar}.

As expected, the holographic results \eqref{S_inv} and \eqref{W_tensor}-\eqref{W_scalar} show that the generating functional \eqref{W_Sinv_Wt} satisfies the KMS conditions, which give rise to the familiar relations among the correlators in \eqref{G_RAS}
\begin{align}
G_{\r\a}^{\mu\nu|\rho\sigma}(K) = \left[ G_{\a\r}^{\mu\nu|\rho\sigma}(K) \right]^*, \qquad \qquad G_{\r\r}^{\mu\nu|\rho\sigma}(K) = \frac{2}{\beta_0\omega} {\rm Im} \left[ G_{\r\a}^{\mu\nu|\rho\sigma}(K) \right]
\end{align}
where hydrodynamic limit is assumed. Therefore, we will focus on the results for the retarded correlators $G_{\r\a}^{\mu\nu|\rho\sigma}(K)$, which are classified into three independent channels \cite{Policastro:2002tn}.

For the tensor channel, the retarded correlator is
\begin{align}
	G^{yz|yz}_{\r\a} = r_h^4 \left( \frac{1}{2} -  {\rm i} \tomega + \cdots \right)
\end{align}
which is simply analytic, in agreement with the fact that the tensor channel does not contain any hydrodynamic mode.

For the shear channel, the retarded correlators are
\begin{align}
	G^{0y|0y}_{\r\a} &= r_h^4 \left[\frac{6 {\rm i} \tomega + \tk^2}{2 (2 {\rm i} \tomega - \tk^2)}  + \cdots \right] \nonumber \\
	G^{0y|xy}_{\r\a} &= r_h^4 \left[\frac{2  \tomega \tk}{2 {\rm i} \tomega - \tk^2}  + \cdots \right] \nonumber \\
	G^{xy|xy}_{\r\a} &= r_h^4 \left(\frac{1}{2} + \cdots \right)
\end{align}
which show a shear pole at $\tomega = -{\rm i} \tk^2/2$. This reflects that the shear channel contains a diffusive mode associated with conservation of transverse momentum.

For the sound channel, the retarded correlators involve more components
\begin{align}
	G^{00|00}_{\r\a} =& r_h^4 \left[\frac{3 (5 \tk^2 - 3 \tomega^2)}{8 (3 \tomega^2 - \tk^2)} - {\rm i} \frac{3 \tomega \tk^4}{ (3 \tomega^2 - \tk^2)^2}  + \cdots \right] \nonumber \\
	G^{00|0x}_{\r\a} =& r_h^4 \left[\frac{3 \tk \tomega }{ 3 \tomega^2 - \tk^2} - {\rm i} \frac{6 \tomega^2 \tk^3}{ ( 3\tomega^2 - \tk^2)^2}  + \cdots \right] \nonumber \\
	G^{00|xx}_{\r\a} =& r_h^4 \left[\frac{3  (\tomega^2 + \tk^2) }{ 8 (3 \tomega^2 - \tk^2)} - {\rm i} \frac{3 \tomega^3 \tk^2}{ ( 3\tomega^2 - \tk^2)^2}  + \cdots \right] \nonumber \\
	G^{00|yy}_{\r\a} =& r_h^4 \left[\frac{3  (\tomega^2 + \tk^2) }{ 8 (3 \tomega^2 - \tk^2)} - {\rm i} \frac{3 \tomega \tk^2 (\tk^2 - \tomega^2)}{ 2 ( 3\tomega^2 - \tk^2)^2}  + \cdots \right] \nonumber \\
	G^{0x|0x}_{\r\a} =& r_h^4 \left[\frac{3  (9 \tomega^2 + \tk^2) }{ 2 (3 \tomega^2 - \tk^2)} - {\rm i} \frac{12 \tomega^3 \tk^2}{ ( 3\tomega^2 - \tk^2)^2}  + \cdots \right] \nonumber \\
	G^{0x|xx}_{\r\a} =& r_h^4 \left[\frac{  \tk \tomega }{ 3 \tomega^2 - \tk^2} - {\rm i} \frac{6 \tomega^4 \tk}{ ( 3\tomega^2 - \tk^2)^2}  + \cdots \right] \nonumber \\
	G^{0x|yy}_{\r\a} =& r_h^4 \left[\frac{  \tk \tomega }{ 3 \tomega^2 - \tk^2} - {\rm i} \frac{3 \tomega^2 \tk (\tk^2 - \tomega^2)}{ ( 3\tomega^2 - \tk^2)^2}  + \cdots \right] \nonumber \\
	G^{xx|xx}_{\r\a} =& r_h^4 \left[\frac{ (7 \tomega^2 - \tk^2) }{ 8 (3 \tomega^2 - \tk^2)} - {\rm i} \frac{3 \tomega^5 }{ ( 3\tomega^2 - \tk^2)^2}  + \cdots \right] \nonumber \\
	G^{xx|yy}_{\r\a} =& r_h^4 \left[\frac{ (\tomega^2 + \tk^2) }{ 8 (3 \tomega^2 - \tk^2)} - {\rm i} \frac{3 \tomega^3 (\tk^2 - \tomega^2)}{2 ( 3\tomega^2 - \tk^2)^2}  + \cdots \right] \nonumber \\
	G^{yy|yy}_{\r\a} =& r_h^4 \left[\frac{ (7 \tomega^2 - \tk^2) }{ 8 (3 \tomega^2 - \tk^2)} - {\rm i} \frac{\tomega (3 \tomega^4 -3 k^2 \tomega^2 +  \tk^4) }{ ( 3\tomega^2 - \tk^2)^2}  + \cdots \right] \nonumber \\
	G^{yy|zz}_{\r\a} =& r_h^4 \left[\frac{ (\tomega^2 + \tk^2) }{ 8 (3 \tomega^2 - \tk^2)} - {\rm i} \frac{ \tomega (\tk^4 - 3 \tomega^4)}{2 ( 3\tomega^2 - \tk^2)^2}  + \cdots \right]
\end{align}
Based on the prescription \cite{Son:2002sd} for the Minkowskian space correlators, the leading terms were previously obtained in \cite{Policastro:2002tn}, which predicts a sound pole at $\tomega = \pm \tk/\sqrt{3}$. The sub-leading terms were obtained later in \cite{Arnold:2011ja} using the same prescription. Indeed, it can be shown that \cite{Arnold:2011ja,Crossley:2015evo} the sub-leading terms, combined with the leading ones, produce the attenuation part for the sound mode in \eqref{shear_sound_dispersion}.

\section{Summary and Outlook} \label{summary}

By virtue of the holographic SK contour of \cite{Glorioso:2018mmw}, we derived the SK effective action for a dissipative neutral fluid dual to the Einstein gravity in an asymptotic AdS$_5$ space. This involves the double Dirichlet problem for the linearized gravitational field propagating in a complexified static AdS black brane background. The double AdS boundary conditions for the bulk metric encode the dynamical variables (corresponding to the fluid's velocity and temperature fields \cite{Nickel:2010pr,Crossley:2015tka,deBoer:2015ija}) for writing the boundary fluid's effective action. To first order in the derivatives of the AdS boundary data, we obtained the analytical solutions for the bulk gravitational field based on a partially on-shell scheme. Then, we computed the partially on-shell bulk action to first order in the boundary derivative and to cubic order in the AdS boundary data. Indeed, the partially on-shell bulk action is identified as the boundary fluid's effective action in the fluid spacetime \cite{Crossley:2015evo,Liu:2018kfw}.

We confirmed the holographic effective action by recovering various results known in the context of the classical hydrodynamics. To this end, we have to rewrite the effective action in the physical spacetime, which turned out to be subtle owing to the linearization approximation undertaken throughout this project. Nevertheless, we successfully obtained the hydrodynamic effective action in the physical spacetime, from which we correctly reproduced the hydrodynamic modes, the hydrodynamic constitutive relation \eqref{hydro_const_relation} as well as the full set of two-point correlation functions of the stress tensor.

There are several interesting directions we would like to pursue in the near future.

First, it is of importance to go beyond the linearization approximation for both the bulk gravity and the boundary fluid action. The former can be tackled via the Green's function approach for solving nonlinear corrections for the bulk metric. The latter seems to be more subtle when it comes to rewriting the fluid action in the physical spacetime. Particularly, the boundary data $B_{\s ab}$ would be a highly nonlinear quantity as viewed in the physical spacetime.

Second, it is very interesting to deepen our understanding of the fluid re-parameterization symmetry proposal \eqref{fluid_reparam} from the dual gravity perspective. The recent interesting work \cite{Knysh:2024asf} has initialized such a study, suggesting a link between re-parameterization symmetry for the fluid and the horizon symmetry of the dual gravitational spacetime. A specific question is to understand the frame issues for hydrodynamics, or, more generally, the redefinition of the dynamical fields in the hydrodynamic EFT, from the holographic SK technique. We expect this to be related to the undertaken horizon condition \eqref{horizon_condition} to achieve the fluid re-parameterization symmetry \eqref{fluid_reparam}.

Last but not the least, it would be interesting to take a local equilibrium state along the line of \cite{Bhattacharyya:2008jc,Rangamani:2009xk,Crossley:2015tka} and work out the bulk corrections perturbatively in the boundary derivative expansion. This would up-lift the fluid-gravity correspondence into a hydrodynamic EFT framework with the fluid-gravity correspondence representing the mean-field part of the effective action. Study along this line requires to promote the constant horizon radius $r_h$ of Figure \ref{holographic_SK_contour} into a coordinate-dependent function $r_h(x^\mu)$. Here, the main difficulty would be to find out the ``seed geometry'', i.e., a boundary spacetime dependent AdS geometry corresponding to a local equilibrium state. This direction will overlap with the first direction above on going beyond the linearization approximation.

\appendix

\section{Various terms in \eqref{separation2} and \eqref{tilde_W}} \label{cal_detail}

In the Fourier space, the invariant part $S_{\rm inv}$ in \eqref{separation2} is
\begin{align}
S_{\rm inv} & = \frac{r_h^4}{2} \left( 3 A_{\a 00}(K) + A_{\a xx}(K) \right)\delta^{(4)}(K) - r_h^4 \left(3 A_{\r 0\alpha}(-K) A_{\a 0\alpha}(K) + A_{\r x\alpha}(-K) A_{\a x\alpha}(K) \right) \nonumber \\
& + \frac{3 r_h^4}{4} \left[ A_{\r 00}(-K) A_{a00}(K) - \frac{1}{\omega^2} (k A_{\r 00}(-K) + 2 \omega A_{\r 0x}(-K))(k A_{\a 00}(K) + 2 \omega A_{\a 0x}(K)) \right] \nonumber \\
& -\frac{ r_h^4}{4} \left[ A_{\r xx}(-K) A_{axx}(K) + A_{\r +}(-K) (A_{a00}(K) - A_{\a xx}(K)) + A_{\a +}(-K) (A_{\r 00}(K) - A_{\r xx}(K)) \right.\nonumber \\
& \left. -\frac{1}{k^2} (\omega A_{\r xx}(-K) + 2 k A_{\r 0x}(-K))(\omega A_{\a xx}(K) + 2 k A_{\a 0x}(K)) \right]. \label{S_inv}
\end{align}
In the Fourier space, the second part in \eqref{separation2} is
\begin{align}
\tilde S_{eff} = & \frac{3 k^2 r_h^4}{4 \omega^2} B_{\r 00}(K) B_{\a 00}(-K) - \frac{\omega^2 r_h^4}{4 k^2} B_{\r xx}(K)B_{\a xx}(-K) \nonumber \\
& + \frac{3 k r_h^4}{2 \omega} \left[ B_{\r 00}(K)B_{\a 0x}(-K) + B_{\r 0x}(K) B_{\a 00}(-K) \right] \nonumber \\
&-\frac{\omega r_h^4}{ 2 k}\left[ B_{\r xx}(K)B_{\a 0x}(-K) + B_{\r 0x}(K)B_{\a xx}(-K) \right] \nonumber \\
& -\frac{r_h^4}{4} B_{\r 00}(K)B_{\a xx}(-K) + \frac{3 r_h^4}{4} B_{\r xx}(K)B_{\a 00} (-K) + 3 r_h^4 B_{\r 0x}(K)B_{\a 0x}(-K) \nonumber \\
&+ 4 r_h^4 B_{\r 0 \alpha}(K) B_{\a 0 \alpha}(-K) + r_h^4 B_{\r \alpha \alpha}(K)B_{\a 00}(-K) \nonumber \\
&+ \frac{r_h^4}{4}\left[ B_{\r \alpha \alpha}(K) B_{\a \beta \beta}(-K) - 2 B_{\r \alpha \beta}(K)B_{\a \alpha \beta}(-K) \right] \nonumber \\
& + \frac{{\rm i} r_h^4}{6 \pi} \left[ 3 B_{\a ij}(K) B_{\a ij}(-K) - B_{\a ii}(K) B_{\a jj}(-K) \right] \nonumber \\
& + \frac{r_h^3}{6} \left[ B_{\a ii}(K) \partial_0 B_{\r jj}(-K) - 3 B_{\a ij}(K) \partial_0 B_{\r ij}(-K) \right] - \frac{{\rm i}\pi r_h^3}{8} B_{\a 00}(K) \partial_0 B_{\a ii}(-K) \label{tilde_Seff}
\end{align}
In the above expressions, the indices $\alpha, \beta$ mean transverse coordinates $y,z$. Moreover, we have $A_{\r +} \equiv A_{\r yy} + A_{\r zz}$ and similarly for $A_{\a +}$. Notice that in \eqref{S_inv} and \eqref{tilde_Seff}, we have suppressed the integrals over the four-momentum $K_\mu$, i.e., $\int d^4K/(2\pi)^4$.

The various parts in \eqref{tilde_W} are
\begin{align}
\tilde W_{\rm tensor} =& r_h^4 (-1+2 {\rm i} \tilde \omega )\left[ A_{\r\alpha \beta}(K) A_{\a\alpha \beta}(-K)  + \frac{1}{4}A_{\r -}(K) A_{\a -}(-K)\right]  \nonumber \\
&+ \frac{{\rm i} r_h^4}{\pi } \left[ A_{\a\alpha \beta}(K) A_{\a\alpha \beta}(-K)  + \frac{1}{4} A_{\a -}(K) A_{\a -}(-K) \right], \label{W_tensor}
\end{align}
\begin{align}
\tilde W_{\rm vector} =& \frac{8 r_h^4 {\rm i} \tilde \omega }{\tilde k^2-2 {\rm i} \tilde \omega } A_{\r 0\alpha}(K) A_{\a 0\alpha}(-K) + \frac{4 r_h^4 \tilde k \tilde \omega }{\tilde k^2-2 {\rm i} \tilde \omega } \left[ A_{\r 0\alpha}(K) A_{\a x \alpha}(-K) + A_{\r x\alpha}(K) A_{\a 0\alpha}(-K) \right] \nonumber \\
& + \frac{r_h^4}{\pi}\frac{4 {\rm i} \tilde k^2}{\tilde k^4+4 \tilde \omega ^2} A_{\a 0\alpha}(K) A_{\a 0\alpha}(-K)  + \frac{{\rm i} r_h^4}{\pi}\frac{\tilde k^4 }{ \tilde k^4 + 4\tilde \omega^2} A_{\a x\alpha}(K) A_{\a x\alpha}(-K) \nonumber \\
& + \frac{{\rm i} r_h^4}{\pi} \frac{8 \tilde k \tilde \omega }{\tilde k^4 + 4 \tilde \omega^2}  A_{\a 0\alpha}(K) A_{\a x\alpha}(-K), \label{W_vector}
\end{align}
\begin{align}
\tilde W_{\rm scalar} =& \left[ \frac{3 r_h^4}{\tilde k^2-3 \tilde \omega^2} + \frac{6 {\rm i} r_h^4 \tilde k^2 \tilde \omega }{\left(\tilde k^2 - 3 \tilde \omega ^2\right)^2} \right] \left[ \tilde k^2 A_{\r 00}(K) A_{\a 00}(-K) + 4 \tilde \omega^2 A_{\r 0x}(K) A_{\a 0x}(-K) \right] \nonumber \\
&+ \left[ \frac{6 \tilde k  r_h^4 \tilde \omega }{\tilde k^2-3 \tilde \omega^2} + \frac{12 {\rm i} \tilde k^3 r_h^4 \tilde \omega ^2}{\left(\tilde k^2-3 \tilde \omega^2 \right)^2} \right] \left[ A_{\r 00}(K) A_{\a 0x}(-K) + A_{\r 0x}(K) A_{\a 00}(-K) \right]\nonumber \\
&  + \left[ \frac{2 \tilde k r_h^4 \tilde \omega }{\tilde k^2-3 \tilde \omega^2} + \frac{12 {\rm i} \tilde k r_h^4 \tilde \omega^4}{\left(\tilde k^2-3 \tilde \omega ^2\right)^2} \right] \left[ A_{\r 0x}(K) A_{\a xx}(-K) + A_{\r xx}(K) A_{\a 0x}(-K) \right] \nonumber \\
& + \left[ \frac{\tilde k^2 r_h^4}{\tilde k^2-3 \tilde \omega ^2}+\frac{6 {\rm i} \tilde k^2 r_h^4 \tilde \omega^3}{\left(\tilde k^2-3 \tilde \omega^2\right)^2} \right] A_{\r 00}(K) A_{\a xx}(-K)  \nonumber \\
& + \left[ \frac{3 r_h^4 \tilde \omega^2}{\tilde k^2-3 \tilde \omega^2}+\frac{6 {\rm i} \tilde k^2 r_h^4 \tilde \omega ^3}{\left(\tilde k^2-3 \tilde \omega^2\right)^2} \right] A_{\r xx}(K) A_{\a 00}(-K) \nonumber \\
&+ \left[ \frac{\tilde k^2 r_h^4}{\tilde k^2-3 \tilde \omega^2}+\frac{3 {\rm i} \tilde k^2 r_h^4 \tilde \omega \left(\tilde k^2-\tilde \omega^2\right)}{\left(\tilde k^2-3 \tilde \omega ^2\right)^2} \right] A_{\r 00}(K) A_{\a +}(-K) \nonumber \\
&+ \left[ \frac{3 r_h^4 \tilde \omega ^2}{\tilde k^2-3 \tilde \omega ^2}+\frac{3 {\rm i} \tilde k^2 r_h^4 \tilde \omega \left(\tilde k^2-\tilde \omega ^2\right)}{\left(\tilde k^2-3 \tilde \omega ^2\right)^2} \right] A_{\r +}(K)A_{\a 00}(-K) \nonumber \\
&+ \left[ \frac{2 \tilde k r_h^4 \tilde \omega }{\tilde k^2-3 \tilde \omega^2} + \frac{6 {\rm i} \tilde k r_h^4 \tilde \omega^2 \left(\tilde k^2-\tilde \omega ^2\right)}{\left(\tilde k^2-3 \tilde \omega^2\right)^2} \right] \left[ A_{\r 0x}(K)A_{\a +}(-K) + A_{\r +}(K) A_{\a 0x}(-K) \right] \nonumber \\
&+ \left[ \frac{ r_h^4 \tilde \omega^2}{\tilde k^2-3 \tilde \omega^2}-\frac{2 {\rm i} \tilde k^2 r_h^4 \tilde \omega \left(\tilde k^2-6 \tilde \omega^2\right)}{3 \left(\tilde k^2-3 \tilde \omega ^2\right)^2} \right] A_{\r xx}(K) A_{\a xx}(-K) \nonumber \\
&+ \left[ \frac{ r_h^4 \tilde \omega^2}{\tilde k^2-3 \tilde \omega^2}+\frac{{\rm i} \tilde k^2 r_h^4 \tilde \omega \left(\tilde k^2+3 \tilde \omega^2\right)}{3 \left(\tilde k^2-3 \tilde \omega^2\right)^2} \right] \left[ A_{\r xx}(K) A_{\a +}(-K) + A_{\r +}(K) A_{\a xx}(-K) \right] \nonumber \\
& + \left[\frac{ r_h^4 \tilde \omega^2}{\tilde k^2-3 \tilde \omega^2}+\frac{2 {\rm i} r_h^4 \left(2 \tilde k^4 \tilde \omega -3 \tilde k^2 \tilde \omega^3\right)}{3 \left(\tilde k^2-3 \tilde \omega^2\right)^2} \right] A_{\r +}(K) A_{\a +}(-K) \nonumber \\
& + \frac{3 {\rm i} \tilde k^4 r_h^4}{\pi \left(\tilde k^2-3 \tilde \omega^2\right)^2} A_{\a 00}(K) A_{\a 00}(-K) + \frac{12 {\rm i} \tilde k^3  r_h^4 \tilde \omega }{\pi  \left(\tilde k^2-3 \tilde \omega^2\right)^2} A_{\a 00}(K) A_{\a 0x}(-K) \nonumber \\
& + \frac{12 {\rm i} \tilde k^2  r_h^4 \tilde \omega^2}{\pi \left(\tilde k^2-3 \tilde \omega^2\right)^2} A_{\a 0x}(K) A_{\a 0x}(-K) + \left[ \frac{1}{4} \pi r_h^4 \tilde \omega + \frac{6 {\rm i} \tilde k^2 r_h^4 \tilde \omega^2}{\pi \left(\tilde k^2-3 \tilde \omega^2\right)^2} \right] A_{\a 00}(K) A_{\a xx}(-K) \nonumber \\
&+\left[ \frac{1}{4} \pi r_h^4 \tilde \omega +\frac{3 {\rm i} r_h^4 \left(\tilde k^4-\tilde k^2 \tilde \omega^2\right)}{\pi \left(\tilde k^2-3 \tilde \omega^2\right)^2} \right] A_{\a 00}(K) A_{\a +}(-K) + \frac{12 {\rm i} \tilde k  r_h^4 \tilde \omega^3}{\pi \left(\tilde k^2-3 \tilde \omega^2\right)^2} A_{\a 0x}(K) A_{\a xx}(-K) \nonumber \\
&+ \frac{6 {\rm i} \tilde k r_h^4 \tilde \omega \left(\tilde k^2-\tilde \omega ^2\right)}{\pi \left(\tilde k^2-3 \tilde \omega^2\right)^2} A_{\a 0x}(K) A_{\a +}(-K) - \frac{{\rm i} r_h^4 \left(\tilde k^4-6 \tilde k^2 \tilde \omega^2\right)}{3 \pi \left(\tilde k^2-3 \tilde \omega^2\right)^2} A_{\a xx}(K) A_{\a xx}(-K) \nonumber \\
&+ \frac{{\rm i} r_h^4 \left(3 \tilde k^2 \tilde \omega^2+\tilde k^4\right)}{3 \pi  \left(\tilde k^2-3 \tilde \omega^2\right)^2} A_{\a xx}(K) A_{\a +}(-K) + \frac{{\rm i} r_h^4 \left(2 \tilde k^4-3 \tilde k^2 \tilde \omega^2\right)}{3 \pi \left(\tilde k^2-3 \tilde \omega^2\right)^2} A_{\a +}(K) A_{\a +}(-K) \label{W_scalar}
\end{align}

Here, it is important to stress that the obvious singularities at $\omega=0, k=0$ in \eqref{S_inv} are exactly cancelled by those in \eqref{W_tensor} through \eqref{W_scalar}. Therefore, in the hydrodynamic limit, the only singularities in the generating functional $W$ correspond to the shear mode and the sound mode in \eqref{shear_sound_dispersion}. This is consistent with the field theory analysis \cite{Policastro:2002tn}.

\section*{Acknowledgements}

We would like to thank Michael Lublinsky for insightful discussions related to this project. This work was supported by the National Natural Science Foundation of China (NSFC) under the grant No.~12375044.

\bibliographystyle{utphys}
\bibliography{reference}
\end{document}